\documentclass[aps, nofootinbib, floatfix, amsmath, amssymb, runinaddress, tightenlines, 11pt]{revtex4-1}
\pdfoutput=1

\usepackage{amssymb,amsmath,amsfonts,amsbsy,epsfig,wrapfig,caption,float,slashed,graphicx, pstricks}
\usepackage[singlelinecheck=off,justification=raggedright]{caption}
\usepackage[utf8]{inputenc}

\newcommand{\eq}[2]{\begin{equation}\label{#1}#2 \end{equation}}

\usepackage{xcolor}
\usepackage[utf8]{inputenc}

\newcommand{\mpl}{M_{\rm pl}}
\newcommand{\fnl}{f_{\rm NL}}

\newcommand{\calL}{{\cal L}}
\newcommand{\calP}{{\cal P}}
\newcommand{\calR}{{\cal R}}

\newcommand{\calN}{{\cal N}}

\newcommand{\R}{{\widehat{\cal R}}}

\def\bea{\begin{eqnarray}}
\def\eea{\end{eqnarray}}
\def\be{\begin{equation}}
\def\ee{\end{equation}}
\def\ba{\begin{array}}
\def\ea{\end{array}}

\def\nn{\nonumber}

\def\nusr{N_{\rm USR}}

\def\be{\begin{equation}}
\def\ee{\end{equation}}
\def\bea{\begin{eqnarray}}
\def\eea{\end{eqnarray}}

\def\fnl{f_{\rm NL}}

\def\mpc{{\rm Mpc}}

\def\R{{\mathcal R}}
\def\P{{\mathcal P}}
\def\Mpc{{\rm Mpc}}
\def\kusr{k_{\rm u}}

\newcommand{\HubMas}{M_\text{H}}

\newcommand{\Mhub}{\HubMas}

\begin{document}
\author{Christian T. Byrnes$^{1}$}
\email{C.Byrnes@sussex.ac.uk}
\author{Philippa S. Cole$^{1}$}
\email{P.Cole@sussex.ac.uk}
\author{Subodh P. Patil$^{2}$}
\email{patil@nbi.ku.dk}
\affiliation{\\ 1) Department of Physics and Astronomy, University of Sussex, Brighton BN1 9QH, United Kingdom\\}

\affiliation{\\2) Niels Bohr International Academy and Discovery Center,\\ Niels Bohr Institute, Blegdamsvej 17,\\ Copenhagen, DK 2100, Denmark\\}
\date{\today}
\title{Steepest growth of the power spectrum and primordial black holes}
\begin{abstract}
We derive analytic bounds on the shape of the primordial power spectrum in the context of single-field inflation. In particular, the steepest possible growth has a spectral index of $n_s - 1 = 4$ once transients have died down. Its primary implication is that any constraint on the power spectrum at a particular scale can be extrapolated to an upper bound over an extended range of scales. This is important for models which generate relics due to an enhanced amplitude of the primordial scalar perturbations, such as primordial black holes. In order to generate them, the power spectrum needs to grow many orders of magnitude larger than its observed value on CMB scales -- typically achieved through a phase of ultra slow-roll inflation -- and is thus subject to additional constraints at small scales. We plot all relevant constraints including CMB spectral distortions and gravitational waves sourced by scalar perturbations at second order. We show how this limits the allowed mass of PBHs, especially for the large masses of interest following recent detections by LIGO and prospects for constraining them further with future observations. We show that any transition from approximately constant $\epsilon$ slow-roll inflation to a phase where the power spectrum rapidly rises necessarily implies an intervening dip in power. We also show how to reconstruct a potential that can reproduce an arbitrary time-varying $\epsilon$, offering a complementary perspective on how ultra slow-roll can be achieved. 
\end{abstract}
\maketitle

\tableofcontents

\section{Introduction}
There has been a recent surge of interest in the possibility that primordial black holes (PBHs) might constitute a non-negligible fraction of the dark matter in the universe. This was largely sparked by the question posed in \cite{Bird:2016dcv} (see also \cite{Clesse:2016vqa,Sasaki:2016jop}) -- whether the order ten solar mass black holes observed by LIGO \cite{Abbott:2016blz} could be primordial\footnote{See however \cite{Naselsky:2016lay,Creswell:2017rbh,Creswell:2018tsr} for an alternative interpretation of the LIGO data.}. This is motivated by the fact that there are several hints for the existence of PBHs \cite{Clesse:2017bsw}, for example the progenitor BH spins of the LIGO detections being consistent with zero in most cases \cite{Abbott:2016nhf, Abbott:2017gyy, Abbott:2017oio,Roulet:2018jbe}, which is expected for PBHs formed during radiation domination \cite{Chiba:2017rvs} (but not matter domination \cite{Harada:2017fjm}) and arguably unexpected for astrophysical BHs \cite{Belczynski:2017gds,Piran:2018bbt}. This begs the follow-up question -- if these observed black holes are of primordial origin\footnote{For more about a stellar origin of the detected BHs see e.g.~\cite{Belczynski:2016obo,deMink:2016vkw,Rodriguez:2016avt}.}, how were they produced and what are the implications for inflationary model building?

The idea that black holes could be primordial relics (albeit of non-thermal origin) was first discussed in \cite{Carr:1974nx}. Since then, the possibility that they could be produced through inflationary dynamics has been vigorously investigated, see e.g.~\cite{Ivanov:1994pa,GarciaBellido:1996qt,Bullock:1996at,Kohri:2007qn,Peiris:2008be,Josan:2010cj,Drees:2011hb,Linde:2012bt,Bugaev:2013fya,Kawasaki:2016pql,Germani:2017bcs,Hertzberg:2017dkh,Garcia-Bellido:2017mdw,Gong,Kannike:2017bxn,Motohashi:2017kbs,Ballesteros:2017fsr,Cicoli:2018asa,Ozsoy:2018flq}, and \cite{Sasaki:2018dmp} for a recent review of PBHs in the context of the recent observations by LIGO. In order for primordial black holes to form, the primordial power spectrum has to grow by about seven orders of magnitude above the amplitude of $\mathcal{P}_\mathcal{R}\sim2\times10^{-9}$ observed on CMB scales (we discuss the uncertainties in this estimate in Sec.~\ref{sec:model-building}). Growing to such an amplitude on smaller scales takes time during inflation due to the causality of the underlying background field dynamics, but to date no one has quantified just how quickly the power spectrum can grow. We show that at least assuming canonical single-field inflation, neglecting transients, the power spectrum cannot grow faster than $n_s-1=4$, even allowing for arbitrary and instantaneous changes in the derivatives of the inflaton field potential. That is, the (inverse) length-scale $k$ must change by at least an order of magnitude in order for the power spectrum to grow by four orders of magnitude. This implies that any observational constraint on the allowed amplitude of the power spectrum which is tighter than the required amplitude to generate PBHs on a particular scale can be extended over a broader range of scales than directly implied by the observations, due to the restriction on how quickly the power spectrum can grow. We also discuss how observational constraints from the CMB, large-scale structure, spectral distortions and Pulsar Timing Arrays (PTA) all provide constraints on the allowed masses of PBHs which could have formed. 

In the context of PBHs as dark matter (DM), the PBH mass function is important for determining the fraction of the energy density in DM that could be made up of PBHs given current constraints on their detection. We therefore investigate whether restrictions on the primordial power spectrum growth rate have an effect on the PBH mass function and find that vastly different power spectra produce very similar mass functions. This means that if one is interested in producing PBHs within a particular range of masses, observational constraints on the power spectrum will need to be avoided without the slope increasing faster than $k^4$, and the resulting mass function - which will be largely independent of the power spectrum - must then also avoid constraints on the allowed fraction of PBHs in dark matter. Placing analytic bounds on the steepest growth of the power spectrum defines the largest windows possible for PBH production, and targets for future experiments to aim for. 

In section \ref{fastest_growth} we define the slow-roll approximation and deviations from it, and use analytical approximations to find the steepest growth of the power spectrum, and discuss its possible physical basis. We also discuss a dip in the power spectrum that is common to both numerical and analytical results. Fig.~\ref{fig:different-etas} shows the steepest growth. In section \ref{mass_function} we look at the dependence of the mass function on the shape of the power spectrum. In section \ref{observational_constraints} we review the relevant observational constraints on the power spectrum and discuss the effect our bound on the power spectrum has for model-building. Fig.~\ref{fig:new-bringmann} shows our ``master'' plot of the constraints on the power spectrum across a huge range of scales, and future forecasts are shown in Fig.~\ref{fig:forecast}. Finally, we present our conclusions in section \ref{conc} with various details deferred to the appendices.

\section{The Fastest possible growth in power}\label{fastest_growth}

The simplest models of single-field inflation can be described by the so-called slow-roll approximation. This assumes that the inflaton field's kinetic energy is very sub-dominant compared to the potential as it descends it, and is described by the slow-roll parameters:
\bea \epsilon&=&-\frac{\dot{H}}{H^2}=\frac{\dot{\phi}^2}{2H^2\mpl^2}, \label{epsilon} \\
\label{eta} \eta&=&\frac{\dot{\epsilon}}{\epsilon{H}}. 
\eea
These are the first two terms in the so-called Hubble hierarchy, defined (for $i \geq 1$) as
\eq{}{\epsilon_{i+1}:= \frac{\dot \epsilon_i}{H \epsilon_i} .}
For the background to be inflating, $\epsilon$ must be less than unity, and provided it varies slowly as inflation progresses, the resulting primordial power spectrum is nearly scale-invariant. The Planck collaboration \cite{Akrami:2018vks} have measured the amplitude of the primordial power spectrum at scales sampled by the CMB ($k\sim10^{-3}-10^{-1}\ \Mpc^{-1}$) to be of the order $10^{-9}$ and nearly scale-invariant -- consistent with the simplest models of slow-roll inflation. However, CMB measurements tell us nothing about the power spectrum at scales $k\gg  1\, \Mpc^{-1}$. The tightest constraints for $k\gtrsim1\ \Mpc^{-1}$ are disputed, but are certainly orders of magnitudes weaker than those on CMB scales (however, see \cite{Karami:2018qrl} for recent claims to the contrary). For $k\gtrsim10^7\ \Mpc^{-1}$, the constraint is $\mathcal{P}_\mathcal{R}(k)\lesssim10^{-2}$ \cite{Carr:2009jm}, where we use $\mathcal{P}_\mathcal{R}(k)$ to denote the dimensionless power spectrum of the comoving curvature perturbation. This means that the power spectrum is free to grow to around $10^{-2}$ on small scales, and such growth would indicate strong deviations from the standard slow-roll regime. Any peak-like features in the power spectrum are of topical interest since primordial black holes are produced if the power spectrum is of order $10^{-2}$ \cite{Carr:2009jm}. In what follows, we analytically derive a steepest growth index of $k^4$ for the power spectrum in the context of single-field inflation. 

\subsection{Slow-Roll, Beyond Slow-Roll, Ultra Slow-Roll Inflation}

In order for the power spectrum to grow during single-field inflation, the potential must become very flat, meaning that $\epsilon$ must decrease rapidly. The quantity that governs this is $\eta$, as can be seen from the Klein-Gordon equation for a minimally coupled scalar field: 
\eq{eq:KG}{\ddot{\phi}+3H\dot{\phi}+\frac{{\rm d}V}{{\rm d}\phi}=0.}
When the potential is exactly flat, ${\rm d}V/{\rm d}\phi=0$, so that
\eq{eta-6}{-\frac{\ddot{\phi}}{\dot{\phi}H} = \epsilon - \frac{\eta}{2}  = 3. }
Hence the smallest value of $\eta$ attainable for a monotonically decreasing potential is $\eta = -6$. Through the defining equation (\ref{eta}), we see that the fastest $\epsilon$ can therefore decrease is
\eq{}{\epsilon \propto e^{-6\calN},} 
where $\calN$ is the number of e-folds, and we have used $ \frac{d}{dt} = H\frac{d}{d\calN}$. The limiting case for a monotonic potential is an inflection point or an extended period of $V'=0$. As verified explicitly through a potential reconstruction exercise in App.~\ref{appendix:reconst}, we indeed see that a phase of Ultra Slow-Roll (USR) inflation \cite{Tsamis:2003px,Kinney:2005vj}, defined as a phase of constant $\eta = -6$, is attained as one approaches an inflection point, which can also be reasonably well approximated by a small enough first derivative \cite{Dimopoulos:2017ged, Hertzberg:2017dkh,Garcia-Bellido:2017mdw,Germani:2017bcs,Biagetti:2018pjj}. 

\begin{figure}
\centering
\includegraphics[width=\textwidth]{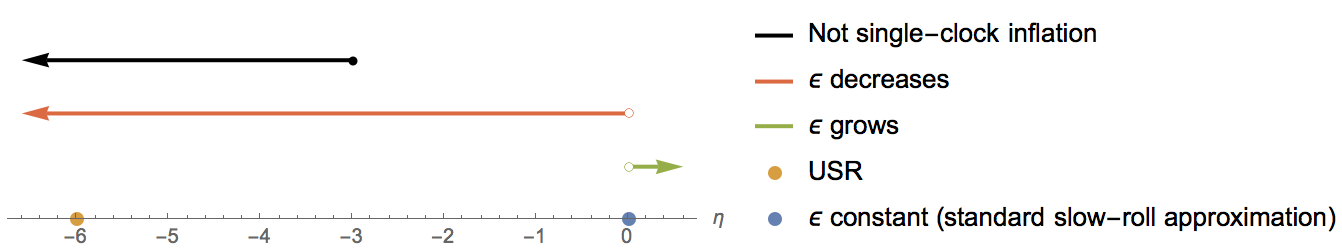}
\caption{The behaviour of the backgrounds and perturbations as a function of the slow-roll parameter $\eta$.}
\label{fig:category}
\end{figure}

If the potential is non-monotonic and the inflaton field rolls uphill then an arbitrarily negative value of $\eta$ is possible, but the potential needs to be extremely tuned and will have many transients associated with it (which could end up dominating) \cite{Martin:2012pe}. For this reason we will mainly focus on regimes of $\eta\geq-6$, however, we will show that our result for the steepest growth of the power spectrum also holds for non-monotonic potentials with $\eta<-6$. Any deviation from $\eta\simeq0$ goes beyond what has typically come to be known as the standard
slow-roll (SR) approximation, which consists of neglecting the acceleration term in \eqref{eq:KG}, a valid approximation only when $\eta\simeq0$. However, qualitatively different behaviour for the mode functions can result from different regimes of $\eta < 0$ even as the background remains approximately de Sitter, with $a(t) \sim e^{H t}$. We see this by first recalling the equation of motion for the curvature perturbation in conformal time
\eq{eq:a3eps}{\calR_k'' + 2\frac{z'}{z}\calR_k' + k^2\calR_k = 0}
where $z^2 = 2a^2\mpl^2\epsilon$. In the long wavelength limit we find the general solution
\eq{}{\calR_{k \to 0} = C_k + D_k\int^\tau\frac{d\tau'}{a^2\epsilon}  .}
The first term is the usual constant super-horizon mode, and the second term ordinarily decays. However, when $\epsilon$ decays at least as fast as $\epsilon \propto a^{-3}$ in cosmological time (i.e.~$a^{-2}$ in conformal time), the second term no longer decays. That is, for $\eta < -3$, one has a growing super-horizon mode in addition to the usual constant mode, whereas for $\eta > -3$ one has the customary constant mode and decaying mode of standard slow-roll inflation. This implies that our inflationary background is not an attractor whenever $\eta \leq -3$, and we are in the peculiar regime of single-field, but non-single-clock inflation. This is because on an attractor, we only have one linearly independent perturbation that can persist -- a local reparameterisation of the background solution, with the other linearly independent perturbation decaying exponentially. When this is no longer the case, an arbitrary perturbation can no longer be described as just a local time reparameterisation of the background -- the defining characteristic of the single-clock regime\footnote{See \cite{Creminelli:2004yq} for a detailed discussion of this point.}. For this reason, we find it useful for the purposes of the following discussion to classify different phases of  $\eta$ as standard slow-roll approximation ($\eta \approx 0$), beyond the slow-roll approximation ($-3 < \eta < 0$), and non-single-clock inflation ($\eta\leq-3$) with the limiting case of ultra slow-roll at $\eta=-6$. See Fig.~\ref{fig:category} for a visual representation of these regimes.

Inflationary potentials which have inflection points or sufficiently flat sections have been studied in e.g.~\cite{Germani:2017bcs,Hertzberg:2017dkh,Garcia-Bellido:2017mdw,Ballesteros:2017fsr,Gong,Cicoli:2018asa}, and are generally found to be severely tuned if one stipulates that a peak be produced in the power spectrum with amplitude of order $10^{-2}$. In what follows, we will show that on the way to such a peak, one cannot increase primordial power arbitrarily fast in $k$-space. Were we to consider a phase of strictly constant $\eta$ evolution, one can straightforwardly derive a steepest possible growth of $\calP_\calR \propto k^3$ (cf.~App.~\ref{section:k3}). However this is too simple an approximation, as any realistic inflationary background must eventually exit such a phase. A more careful multi-phase matching calculation demonstrates a steepest growth of $\calP_\calR \propto k^4$. This implies that the generation of primordial black holes due to peaks in the power spectrum is subject to further model-independent integral constraints from CMB spectral distortions and pulsar-timing array bounds.

\subsection{Regimes of constant $\eta$}\label{section:constant-eta}

As noted above, the power spectrum grows quickly if $\epsilon$ decreases quickly -- a process tracked by the second slow-roll parameter $\eta$. In order to determine the fastest possible growth, we consider regimes where $\eta$ decreases monotonically from 0 to different negative values. Finding the behaviour of the power spectrum for instant transitions between different phases is possible analytically via a matching calculation \cite{Rasanen:2018fom}, and stitching together sufficiently many phases of constant $\eta$ evolution can approximate a smooth transition. For the purposes of deriving a steepest growth index, we note that the growth produced by an instant transition between different phases of constant $\eta$ will be steeper than the growth produced in a smooth transition. 

\begin{figure}
\includegraphics[width=0.9\textwidth]{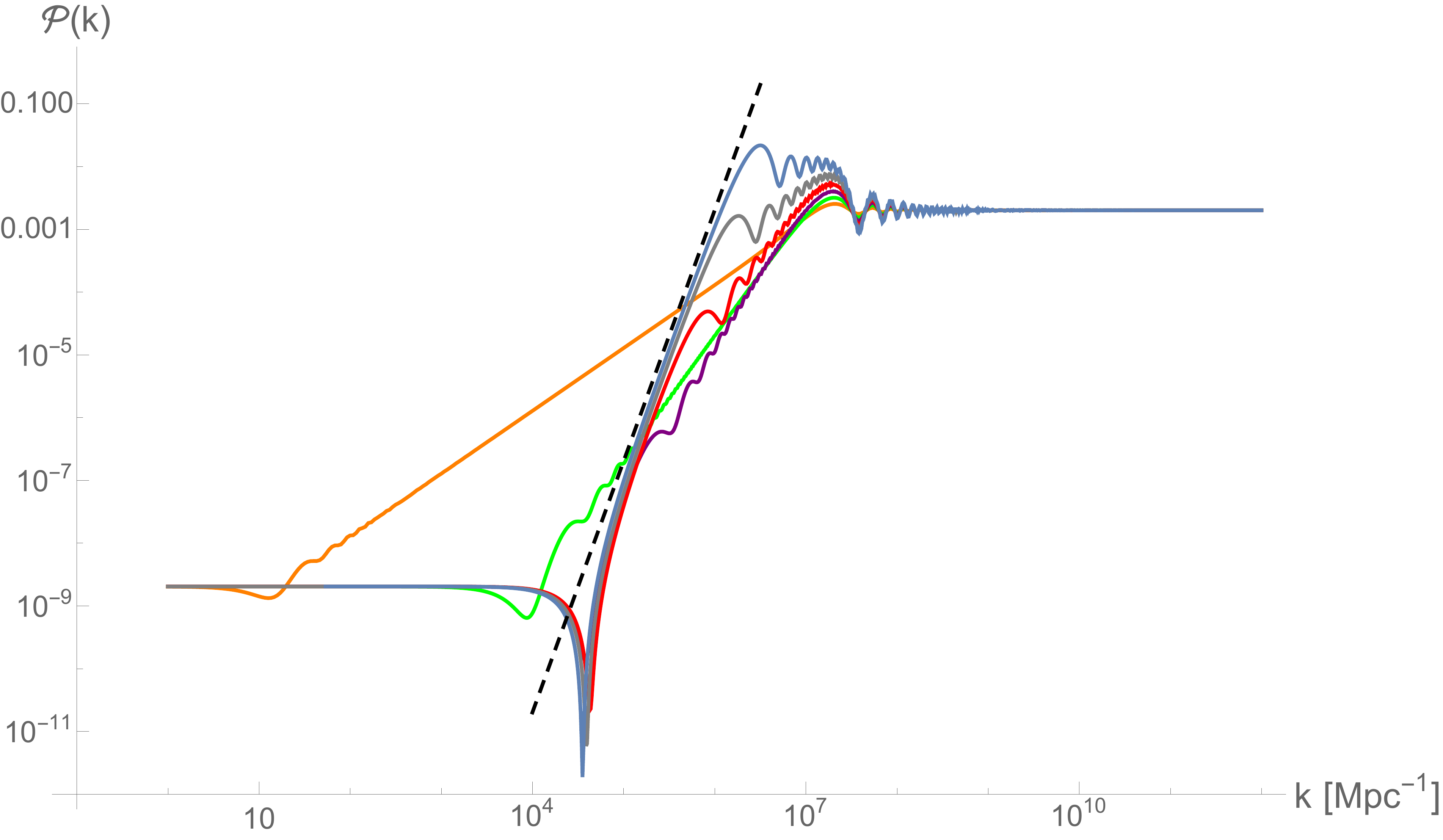}
\caption{Analytical matching from $\eta=0$ to values of constant $\eta$ between -1 and -6, to a final phase of $\eta=0$ slow roll. For $\eta\leq-3$ (purple, red, grey and blue lines in decreasing value of $\eta$), the steepest slope of $n_s-1=4$ is achieved after the dip, before relaxing to a shallower slope decided by the value of $\eta$, see \eqref{eq:ns-constant-eta}. For $\eta>-3$ (green and orange lines), the slope is constant for the whole range of $k$ that the power spectrum increases for. See the online version for colour figures.}
\label{fig:different-etas}
\end{figure}

As elaborated upon in App.~\ref{appendix:matching}, because there are no jumps in the energy momentum tensor of the background between phases of different $\eta$, the Israel junction conditions (see \eqref{jc}) require us to match the curvature perturbation and its first derivatives across the matching surface \cite{Israel:1966rt, Deruelle:1995kd}. All modes begin in the Bunch-Davies vacuum in the initial phase of $\eta=0$, after which we match to a phase of constant $\eta<0$, and then again to a terminal phase of $\eta=0$. We derive analytic expressions for the power spectrum for instant transitions between 0 and the 6 integer values of $\eta$ up to the ultra slow-roll regime in (\ref{srusr}) - (\ref{s2}), the results of which are plotted in Fig.~\ref{fig:different-etas}. The duration of each $\eta < 0$ phase is chosen such that the same growth in amplitude is achieved in all six cases, facilitating a straightforward comparison of the spectral index.

\begin{figure}
\includegraphics[width=0.9\textwidth]{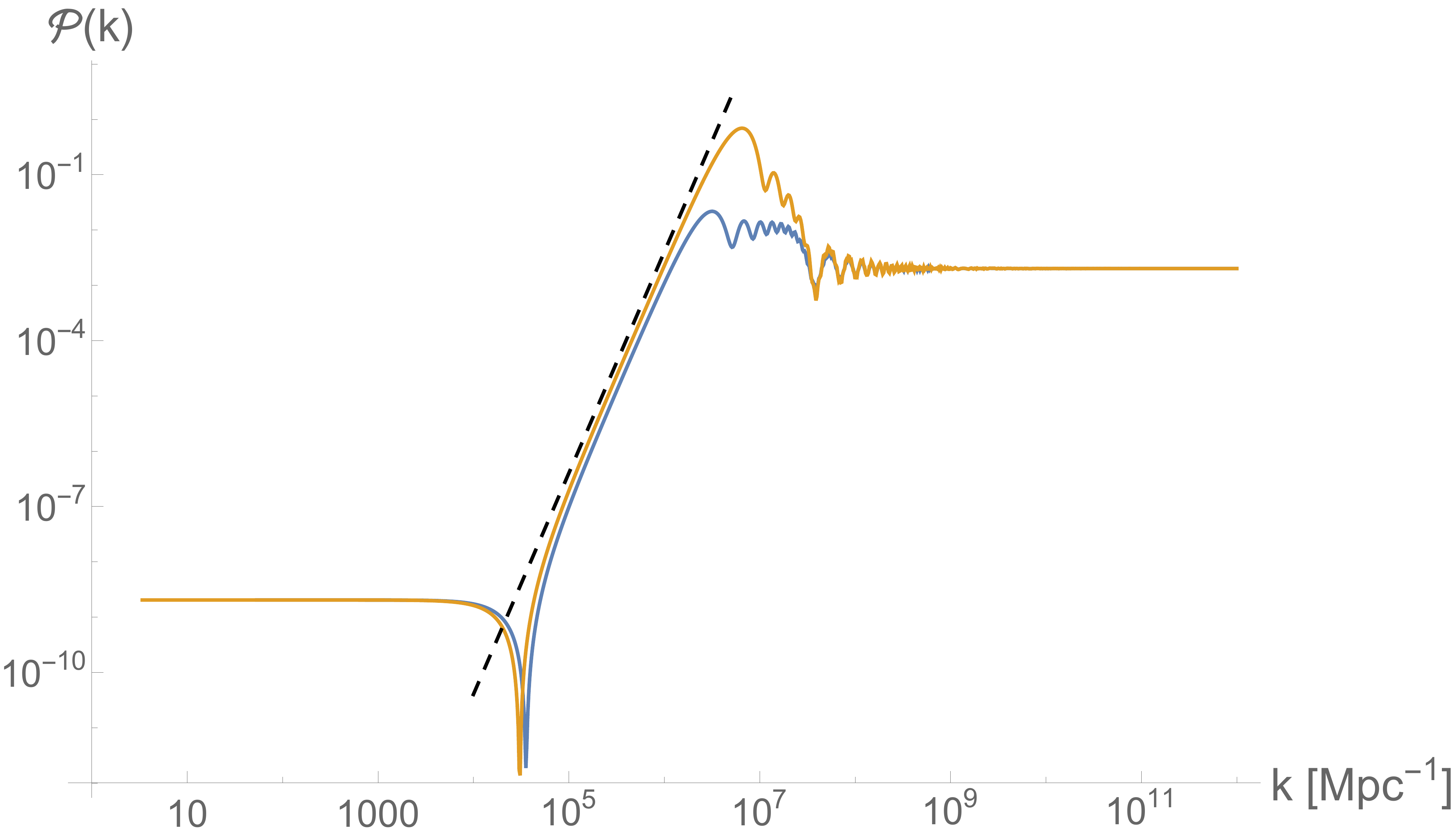}
\caption{The blue line is the same power spectrum as plotted in Fig.~\ref{fig:different-etas} for 2.3 e-folds of USR ($\eta=-6$). The yellow line is a matching from $\eta=0$ to $\eta=-8$ for 1.725 e-folds and back to $\eta=0$. The spectral index of the power spectrum is $n_s=4$ initially (after the dip), followed by a brief period of negative spectral index $n_s\simeq-2$, before scale-invariance for the final $\eta=0$ phase. The dashed black line has a $k^4$ slope.}
\label{fig:eta-8}
\end{figure}

From Fig.~\ref{fig:different-etas}, we see that the slowest growth occurs for the integer values $\eta=-1$, with $n_s-1=1$ once transients have died down (due to the unrealistic instantaneous  transitions\footnote{Note that even smooth but rapid transitions in the inflationary potential can lead to oscillation in the power spectrum, see e.g.~\cite{Adams:2001vc}.}), and for $\eta=-2$, where the growth is given by $n_s-1=2$. Since we only expect the previously decaying mode to start growing once $\eta\leq-3$, the slope of the power spectrum for $\eta>-3$ is determined by the value of $\eta$ only, and matches the expression given in (\ref{eq:ns-constant-eta}). For $\eta\leq-3$, a qualitatively different behaviour emerges. Since we are no longer in the single-clock regime, the previously decaying mode starts to grow, and with it, superhorizon perturbations. Here, the power spectrum has a pronounced dip occurring at scales that exit the horizon a few e-folds before the time of the first transition, followed by an initial growth index proportional to $k^4$, after which it settles to the constant-roll growth given in \eqref{eq:ns-constant-eta}. The initial phase of $k^4$ growth is the steepest possible. In all cases of $\eta\leq-3$, the power spectrum begins to grow before the transition time, which is evidence for superhorizon growth\footnote{In Fig.~\ref{fig:different-etas} the horizon exit scales at the transition times from $\eta=0$ to constant $\eta$ are (for decreasing values of $\eta$ from -1 to -6) $k\sim 10, 10^4, 10^5, 3\times10^5, 6\times10^5 \;{\rm and} \;10^6\, \mpc^{-1}$.}. Providing one adjusts the duration of the $\eta\neq0$ phase such that the final amplitude of the power spectrum is always the same, the rapid growth lasts longest for $\eta=-6$ before reaching a scale-invariant spectrum. Evidently, the steepest growth is characteristic of the non-single clock phase. For an inflationary potential where the inflaton field transiently rolls uphill, $\eta$ can become arbitrarily negative. However one finds a steepest growth of $k^4$ in this case as well,  demonstrated in Fig.~\ref{fig:eta-8} with a matching from $\eta=0$ to $\eta=-8$ and back. We offer an analytic understanding of this steepest growth in the next subsection.

\begin{figure}
\begin{tabular}{c}
\includegraphics[width=0.45\textwidth]{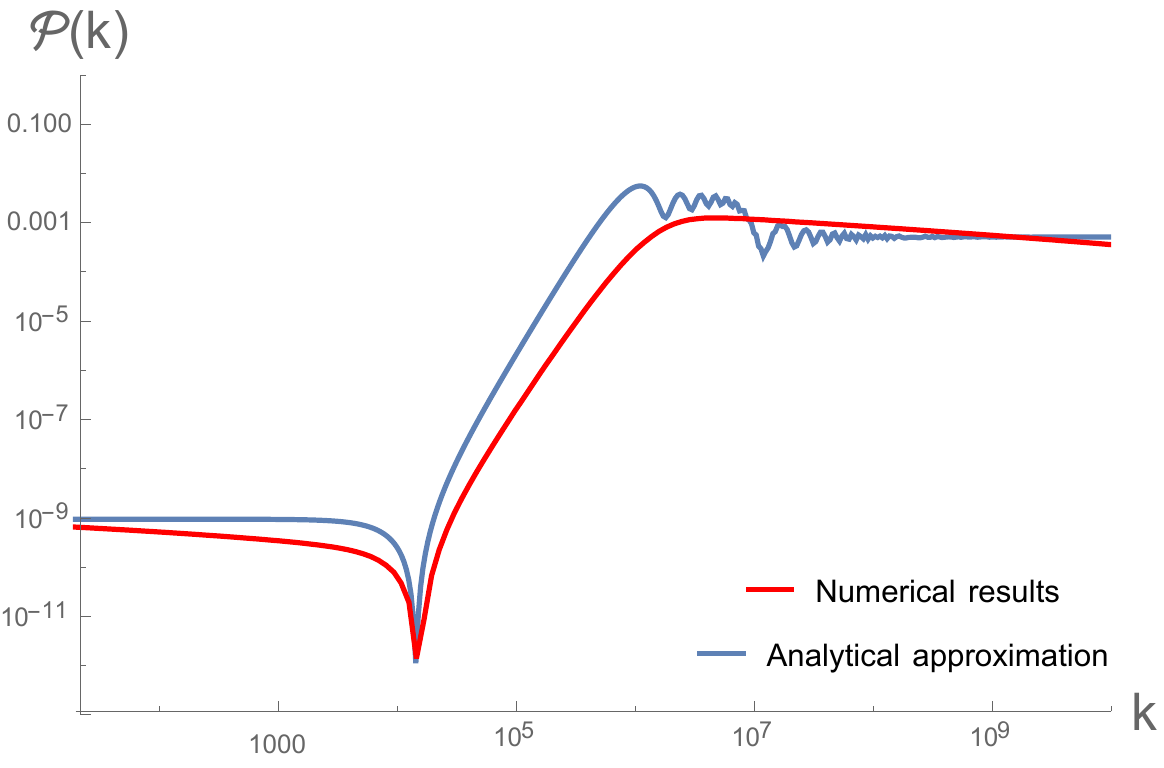}
\hspace{0.5cm}
\includegraphics[width=0.53\textwidth]{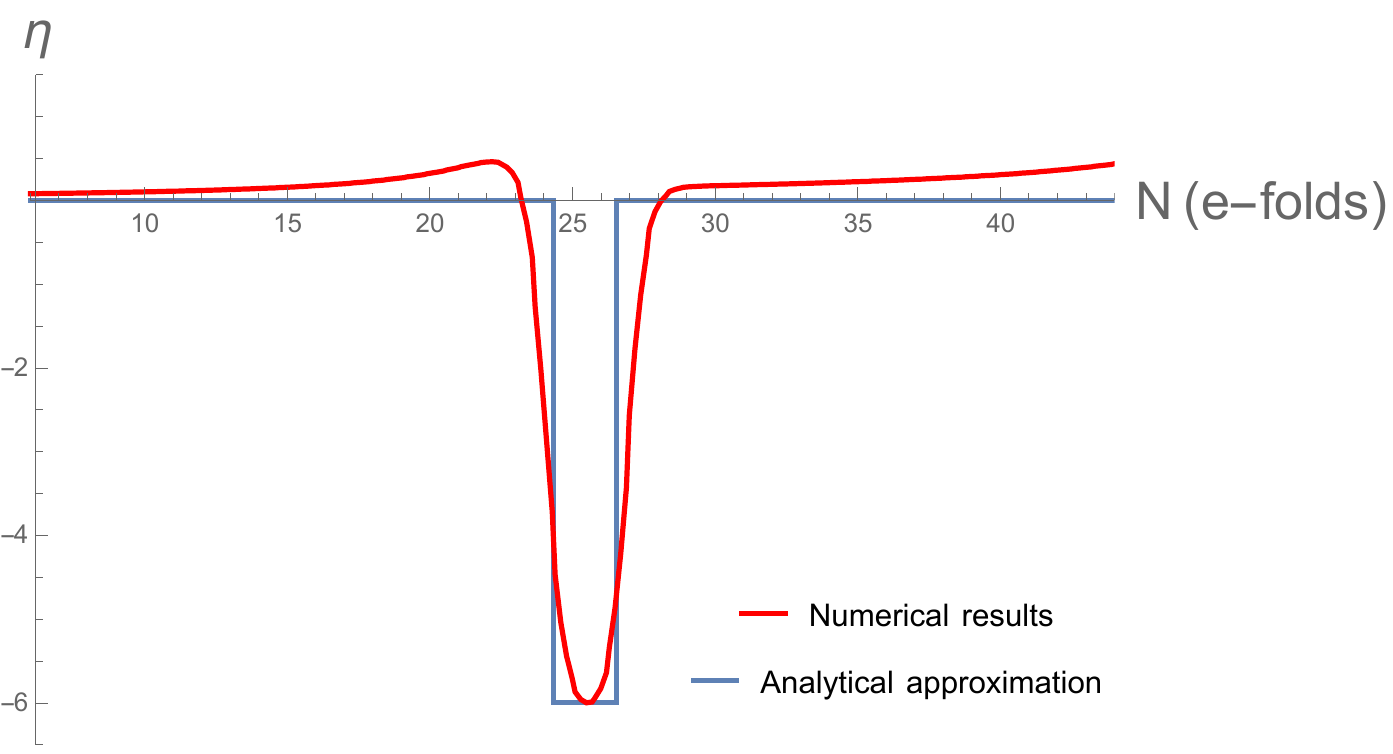}
\end{tabular}
\caption{Left-hand plot: Numerical results for the potential in \cite{Germani:2017bcs} are plotted in red and our analytical approximation is plotted in blue. The analytical approximation involves 3 constant phases of $\eta$ from 0 to -6 and back to 0. The right-hand plot shows the piecewise form for $\eta$ used for the analytical approximation in blue, with 2.2 e-folds of $\eta=-6$. The full numerical evolution of $\eta$ for the potential in \cite{Germani:2017bcs} is shown in red. Note that the units in e-folds have been defined arbitrarily, and we have chosen to centre the phase of $\eta=-6$ in our analytical approximation at the time $N$ when the numerical $\eta$ reaches -6 instantaneously.}
\label{fig:an_num_ps}
\end{figure}

A more realistic treatment would model the evolution of $\eta$ as a series of non-zero constant phases of $\eta$, with instant transitions between each phase to approximate a smooth transition between slow roll and ultra slow roll. As expected, we again find a steepest growth of $k^4$, illustrated in Fig.~\ref{fig:4_matchings}. We can test a `realistic' example of a smooth transition by numerically calculating the final power spectrum for the inflection point model for the potential given in \cite{Garcia-Bellido:2017mdw,Germani:2017bcs}, with the choice of parameters given in section 4 of \cite{Germani:2017bcs}. We use CPPTransport \cite{Seery:2016lko,Dias:2016rjq} to perform the numerical calculation. The red line in Fig.~\ref{fig:an_num_ps} is the resulting numerical power spectrum, and the blue line is the analytical result from approximating the evolution of $\eta$ in the way shown. As expected, the analytic approximation following from an instant matching between phases of inflation grows more steeply than the more realistic (and smooth) potential of \cite{Germani:2017bcs}. 

In App.~\ref{appendix:reconst}, we show how these analytical power spectra might be realised by constructing an example potential that smoothly traverses between $\eta=0$ and $\eta=-6$ and back.

\subsection{Steepest growth and the prior dip in the power spectrum}

Inflationary models which include a phase of non-single-clock evolution (i.e.~with $\eta<-3$) manifest a significant dip in the power spectrum before a steep rise caused by the growth of the perturbations on super-horizon scales, see Fig.~\ref{fig:different-etas}. This has been observed in many recent studies \cite{Hertzberg:2017dkh,Garcia-Bellido:2017mdw,Germani:2017bcs,Biagetti:2018pjj,Cicoli:2018asa,Garcia-Bellido:2016dkw} which have numerically computed the primordial power spectrum for inflationary models with deviations from the slow-roll approximation. It might be assumed that this is caused solely by an increase in $\epsilon$ before the rapid decrease, for example see $\epsilon$ plotted in Fig.~2 of \cite{Cicoli:2018asa}. However, on comparing numerical results with the analytical approximations from section \ref{section:constant-eta} for which $\epsilon$ never increases, we show that the dip cannot be caused by this alone, and that it is actually a generic feature of transitioning from $\eta=0$ to a regime where the decaying mode starts to grow. Perhaps surprisingly, the dip is located on scales which exit the horizon while normal slow roll is still taking place. We explain these features for the particular case of a transition from $\eta=0$ to $\eta=-6$ by making an analytical matching between the two periods. We assume USR lasts for well over an e-folding and neglect the effects of transitioning out of USR, which is a subdominant effect (see Fig.~\ref{fig:peak_power}).

The expansion of the power spectrum in terms of $k/\kusr$, where $\kusr$ is the horizon scale at the time when USR begins (see eq.~(\ref{srusr}) for the full expression), is
\bea \frac{\calP_\calR(k)}{\calP_\calR(0)}= \left[1- \frac45\left(\frac{k}{\kusr} \right)^2 e^{3\nusr} \right]^2\hspace{-5pt}+2\left(\frac{k}{\kusr}\right)^2\hspace{-5pt} -0.10 \left(\frac{k}{\kusr}\right)^6 e^{6\nusr} + 0.0075\left(\frac{k}{\kusr}\right)^8 e^{6\nusr}+\cdots \label{eq:taylor} \eea
where $\nusr$ is the number of e-folds during which $\eta=-6$ and we have dropped terms subleading in $e^{\nusr}$ \footnote{We note that if USR ends in a different way than an instant transition to constant $\epsilon$ then the numerical coefficients in the equation above change slightly, but the qualitative picture remains the same.}. All higher-order terms, which come in even-powers of $k/\kusr$ also come dressed with pre-factors of $e^{6\nusr}$  but with numerical coefficients that are down by an order of magnitude for each even order. Therefore, for $k \lesssim \kusr$, terms up to quartic order are an accurate approximation to the power spectrum. Once $k^2/\kusr^2 \sim \mathcal O (10)$, all of the terms in the alternating series are as important as each other and this is when the series begins to conditionally converge to an oscillating function. Solving for $k$ such that the term in square brackets is zero gives the position of the dip, $k_{\rm dip}$, as
\bea \frac{k_{\rm dip}}{\kusr}=\sqrt{\frac54} e^{-\frac32\nusr} \eea
and hence the dip occurs approximately $\frac32\nusr$ e-folds before USR begins. The amount by which the power spectrum is suppressed at this point is 
\bea \frac{\calP_\calR(k_{\rm dip})}{\calP_\calR(0)}\simeq 2.5 e^{-3\nusr}. \eea
Finally the rapid $k^4$ growth during the transition to USR will end when the $k^4$ and $k^6$ terms in \eqref{eq:taylor} become comparable, which happens when
\bea k\simeq 2.5 \kusr, \eea
and hence occurs about one e-folding after USR has begun, independently of the duration of USR (provided $\nusr\gtrsim1$). 

Thus far, we've arrived at an analytic understanding of the shape of the primordial power spectrum via a matching calculation, and in particular, its steepest possible growth over an intermediate range of scales. This begs the immediate question -- what is the underlying physical mechanism responsible for this steepest growth? Several independent arguments demonstrate a steepest growth of $\calP_\calR \propto k^3$ under the assumption that the large scale power spectrum is a strict power law over all relevant scales. Peebles showed that if the matter power spectrum is to accurately describe particulate matter over scales of cosmological interest, then the two-point function for the density contrast $\delta := \delta\rho/\rho$ can grow no faster than $k^4$ \cite{Peebles}. This implies that the dimensionless power spectrum for the curvature perturbations can grow no faster than $k^3$ since $\partial^2\calR \propto \delta$. As shown in App.~\ref{section:k3}, one can also derive a similar strongest possible scaling for the two point function of the curvature perturbation from the asymptotics of the mode functions. However, none of these arguments apply in the present context, where we do not assume constant power law behaviour for the primordial power spectrum, and the steepest growth is only over a limited (and in principle tunable) interval\footnote{We also note in App.~\ref{section:k3} that in assuming a (possibly distributional) power spectrum of the form $\calP \propto k^n$ at all scales, one can show that it is not possible to regulate the short distance divergence of a spectrum with index $n > 4$ in four dimensions.}. Although one might suspect causality or unitarity arguments to be at play -- or perhaps conformal symmetry as the system tends towards $\epsilon \to 0$ -- it seems that the bound may be due to an interplay of causality arguments and energy-momentum conservation, something we're currently investigating with a particular view to generalising to the multi-field context.

\section{The PBH mass function }\label{mass_function}

Having shown that there are limits to how quickly the power spectrum can grow, one may expect that this also places a sharp limit on how narrow the mass function of PBHs can be. In practice this is not the case, for (at least) three reasons: 1) for any given horizon mass, PBHs form with a spread of comparable masses; 2) the matter power spectrum is less `sharp' than the primordial power spectrum because of the window function relating the two; and 3) PBH formation is exponentially sensitive to the amplitude of the power spectrum, so only perturbations comparable to the peak amplitude are important.

The phenomena of critical collapse \cite{Yokoyama:1998xd,Kuhnel:2015vtw,Carr:2016drx} describes how PBHs of mass $M$ can form with a variety of masses for any given horizon mass $M_H$ according to the relation
\begin{equation}
M=k \Mhub (\delta-\delta_c)^\gamma,
\end{equation} 
where during radiation domination the constants have been numerically estimated as $k=3.3,\;\gamma=0.36,\;\delta_c=0.45$ (the exact values depend upon the radial profile of the perturbations being considered but we use the values given here in order to be concrete) \cite{Niemeyer:1997mt,Musco:2004ak,Musco:2008hv,Musco:2012au}. From the expression given in \cite{Byrnes:2018clq}, the mass function of PBHs, $f(M)$, is 
\begin{eqnarray}
f(M)&\equiv &\frac{1}{\Omega_{\rm CDM}} \frac{d \Omega_{\rm PBH}}{d\ln M}  \nonumber \\ 
&=& \frac{1}{\Omega_{\rm CDM}} \int\limits_{-\infty}^{\infty} \frac{2}{\sqrt{2\pi\sigma^2(\Mhub)}} 
\exp\left[{-\frac{(\mu^{1/\gamma}+\delta_c(\Mhub))^2}{2\sigma^2(\Mhub)}} \right]
\frac{M}{\gamma \Mhub}\mu^{1/\gamma}\sqrt{\frac{M_{\rm eq}}{\Mhub}} d\ln \Mhub,
\end{eqnarray}
where $\mu\equiv\frac{M}{k \Mhub}$.

Inspired by the observation that we cannot have an arbitrarily rapidly growing power spectrum, we calculate the resulting mass function $f(M)$ from 4 different power spectra. The first three grow at different rates towards smaller scales ($n_s-1=0.1,\; 0.2$ and 4 but then drop to zero), the fourth is a Dirac delta function and we chose an overall normalisation of $A_s=0.15$
\bea \P_\calR(k)&=&{\rm 0.242 A_s (k/k_*)^{0.1}  \;{\rm for}\; k<1.5 k_*},\; 0\;\rm{otherwise;} \\
\P_\calR(k)&=&{\rm 0.256  A_s (k/k_*)^{0.2}  \;{\rm for}\; k<1.35 k_*} ,\; 0\;\rm{otherwise;} \\
\P_\calR(k)&=&{\rm  A_s (k/k_*)^{4}  \;{\rm for}\; k<k_*} ,\; 0\;\rm{otherwise;} \\
\P_\calR(k)&=& 0.182 A_s \delta(k-0.83 k_*).
\label{eq:P-unsmoothed}
\eea

The prefactors to the power spectra and the scale at which they drop to zero has been tuned in order to make the position and amplitude of the peak of the mass function as similar as possible, in order to easily compare the width of the mass function. 
The power spectrum dropping instantaneously to zero is unrealistic but unlike an increasing power spectrum, which cannot grow arbitrarily quickly, there is no theoretical limit to how rapidly the spectrum can decay. If the potential is discontinuous and drops to zero instantaneously, then once it goes to zero $\epsilon$ becomes 3 instantly, inflation ends and the universe enters kination domination. For potentials which rapidly switch to a steep negative gradient an arbitrarily rapid transition to a rapidly growing $\epsilon$ can be engineered, which can make the power spectrum as blue as required without ending inflation quickly. For more details, see App.~\ref{sec:end}.

\begin{figure}[H]
	\begin{tabular}{c}
		\includegraphics[scale=0.5]{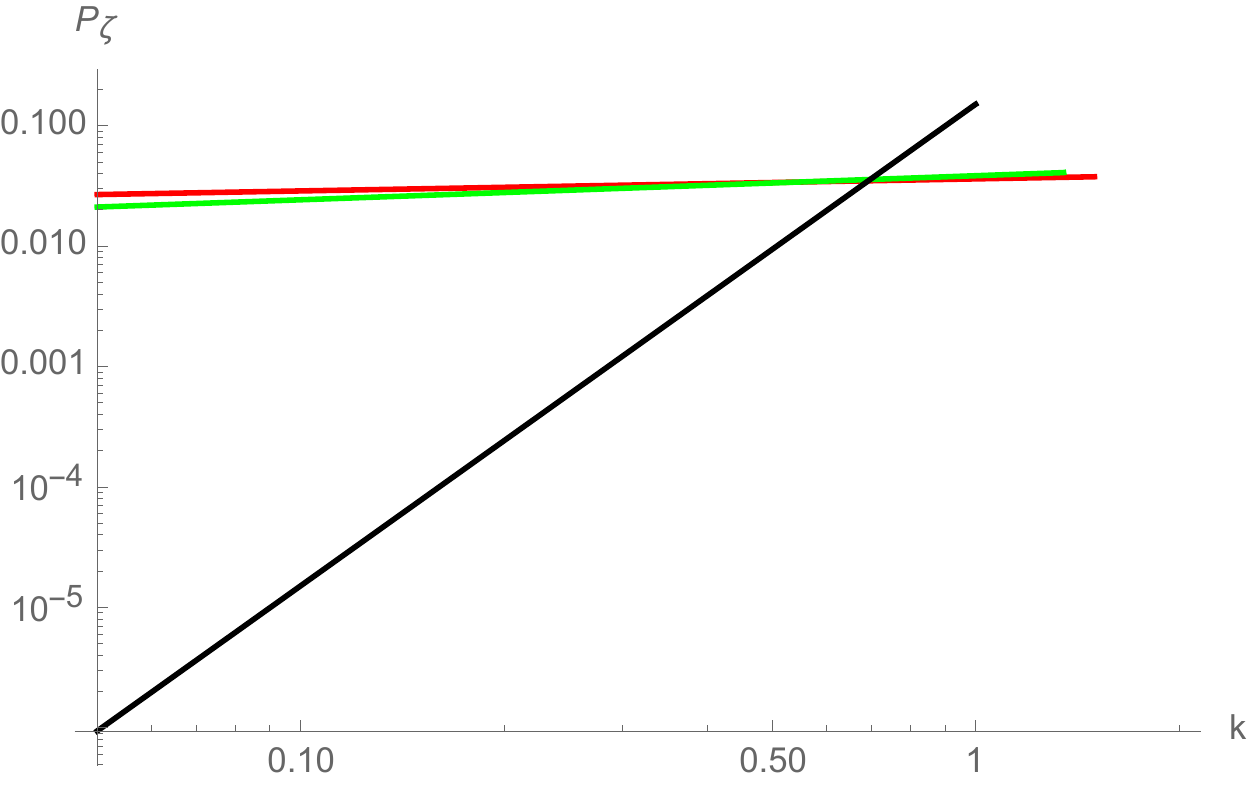}
		\includegraphics[scale=0.47]{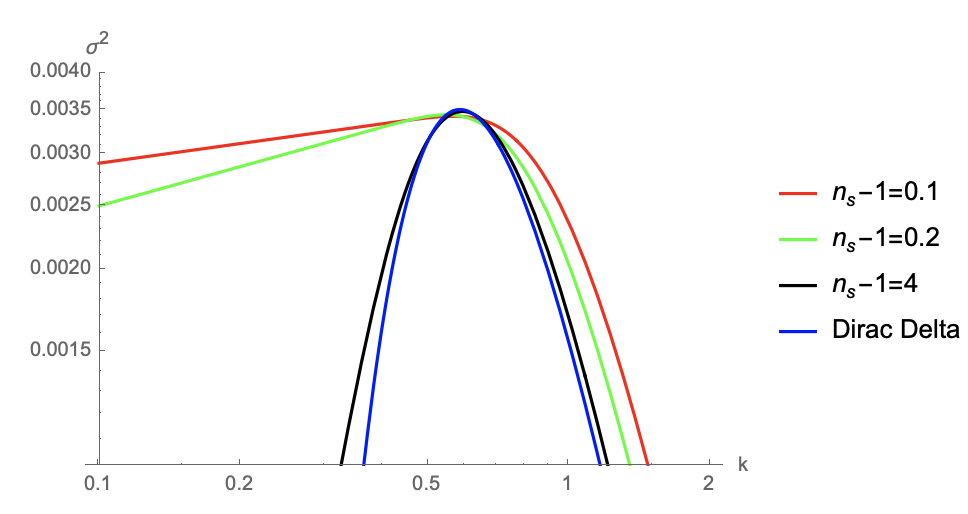}
	\end{tabular}
	\caption{The power spectrum on the left and the smoothed variance of the density contrast on the right for the four models described in the text (the Delta Dirac model is not plotted on the left plot). The power spectra are zero where no line is shown. The x--axis units are arbitrary.}
	\label{fig:P-smoothed}
\end{figure}

\begin{wrapfigure}{H!}{0.48\textwidth}
	\vspace{-3em}
	\begin{center}
		\includegraphics[width=0.45\textwidth, height=3in]{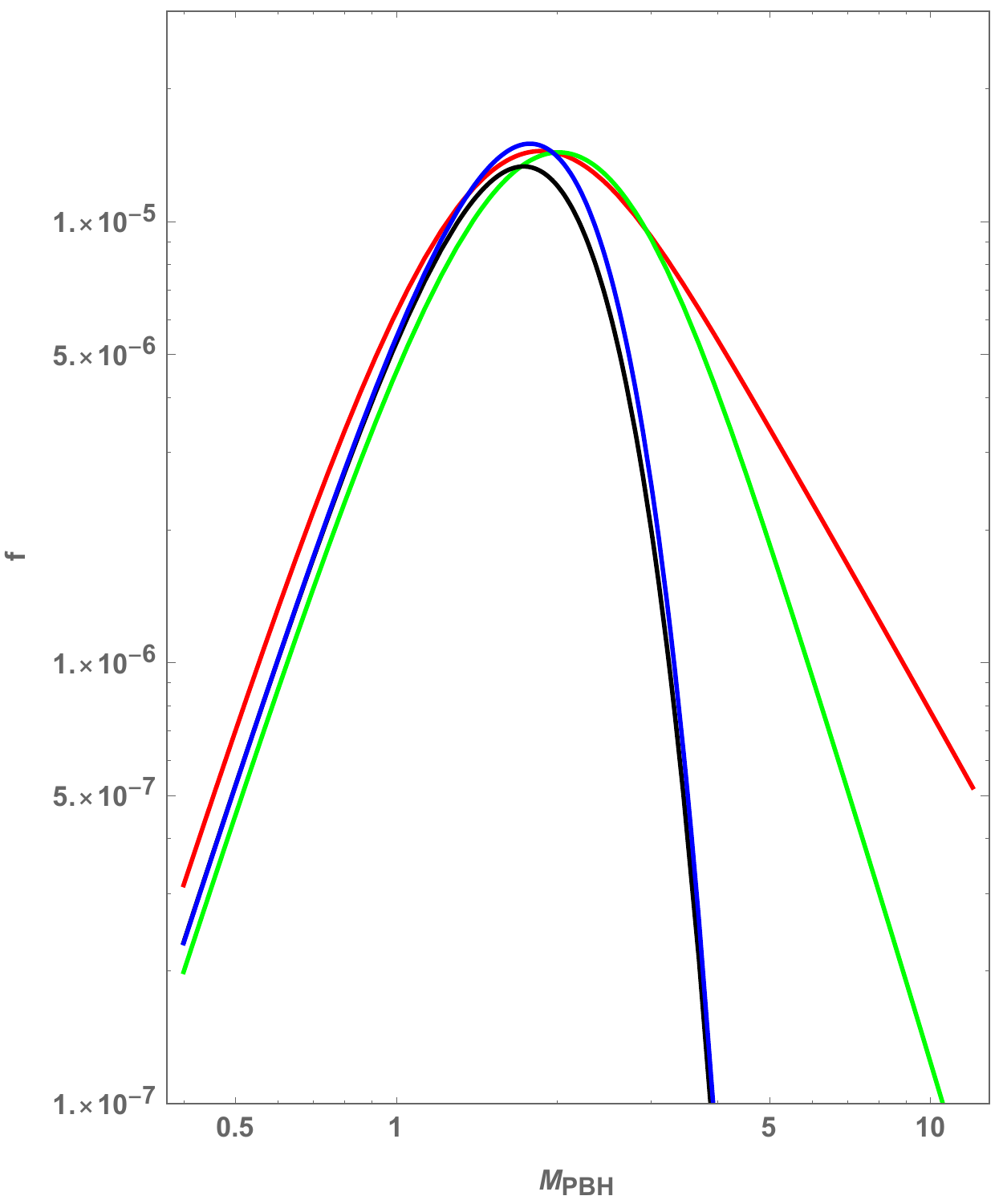}
		\caption{The mass function for the 4 power spectra plotted in Fig.~\ref{fig:P-smoothed}, plotted using the same colour scheme. The arbitrary x--axis units are chosen such that the horizon mass is unity for $k=1$ in the units used for Fig.~\ref{fig:P-smoothed}. 
		\vspace{-2em}
		}
		\label{fig:f-smoothed}		
	\end{center}
\end{wrapfigure}

The variance of the comoving density contrast at horizon entry (smoothed on a scale $R=1/k$) is related to the power spectrum by 
\bea \sigma_R^2
=\int^\infty_0 \frac{{\mathrm d} q}{q} \frac{16}{81}\left(qk^{-1}\right)^4\P_\calR(q) W_R(q)^2
\eea
and we use a Gaussian window function\footnote{The influence of the choice of the window function is discussed in \cite{Ando:2018qdb}. We neglect a transfer function which suppresses sub-horizon perturbations, because it has a negligible impact when using a Gaussian smoothing function.}, $W_R(k)^2=e^{-(kR)^2}$. The results are shown in Fig.~\ref{fig:P-smoothed}.
Fig.~\ref{fig:f-smoothed} demonstrates that the mass function is not very sensitive to how steep or spiked the primordial power spectrum was unless it varied very slowly with scale. There is almost no visible difference in $f(M)$ between a spike which is modelled by a delta Dirac power spectrum or one growing like $k^4$. More surprisingly, even a slowly changing power spectrum with spectral index $n_s-1=0.1$ and a cut off generates a mass function which is not substantially broader than the tightest possible mass function near the peak; compare the black and red lines in Fig.~\ref{fig:f-smoothed}. Note that the mass functions agree extremely well for small masses because the power spectra all have a cut-off scale which has been chosen to align the peaks of the mass functions. The insensitivity of the PBH mass function to the shape of the power spectrum, due to the degeneracies between the effect of the amplitude and shape of the power spectrum, mean that the PBH mass function would have to be detected with very high precision in order to reconstruct the shape of the primordial power spectrum near the corresponding peak.

\section{Observational constraints}\label{observational_constraints}

Although the primordial power spectrum is tightly constrained on CMB scales (from roughly $k \sim 10^{-4} - 10^{-1}$ Mpc$^{-1}$) to be of the order $\calP_\calR \sim 10^{-9}$, there are upper bounds on scales far beyond those accessible in the CMB through a variety of tracers and indirect probes. Of these, the constraints most relevant to the present discussion are distortions of the CMB spectrum from the dissipation of acoustic modes, and bounds from gravitational wave backgrounds produced by scalar perturbations at second order. At the end of this section we produce a ``master plot'' of the key constraints on the power spectrum.

In order to quantify the effect of observational constraints on the allowed number of PBHs generated, which is quantified by $\beta=\rho_{\rm PBH}/\rho_{\rm tot}$ at the time of formation, we need to know the relationship between the amplitude of the power spectrum and $\beta$. In this paper we have neglected the impact of quantum diffusion of the inflaton field and non-Gaussianity of the primordial perturbations. In the context of PBH formation, quantum diffusion during inflation has been discussed subject to the slow-roll approximation by \cite{Pattison:2017mbe} and during USR with conflicting conclusions about its importance in \cite{Biagetti:2018pjj,Ezquiaga:2018gbw,Cruces:2018cvq,Firouzjahi:2018vet}.  

We have also neglected the impact of any non-Gaussianity of the primordial perturbations. This is despite the fact that one of the main reasons that USR inflation was initially considered interesting was that it appeared to generate local non-Gaussianity of order-unity amplitude \cite{Namjoo:2012aa}, providing an exception to the statement that single-field inflation generates negligible local non-Gaussianity \cite{Maldacena:2002vr,Creminelli:2004yq,Mooij:2015yka}. However, Cai et al.~have recently shown that ending USR with a smooth transition tends to erase the local non-Gaussianity \cite{Cai:2017bxr}. If the local non-Gaussianity is not erased or modified by the way USR ends, it has a value $\fnl=5/2$ for modes which exited the horizon long after USR begins (see \cite{Bravo:2017wyw,Bravo:2017gct} for coordinate-choice issues). $\fnl\sim1$ was shown to have a significant impact on the power spectrum constraints in \cite{Byrnes:2012yx}, while the higher-order non-linearity parameters and mode coupling are also important \cite{Young:2013oia,Young:2014oea,Franciolini:2018vbk}. The impact of non-Gaussianity on PBH formation during inflection point inflation was recently considered in \cite{Atal}.

Other uncertainties in relating the amplitude of the power spectrum to the number of black holes include non-sphericity of the initial density profile \cite{Akrami:2016vrq}, the window function used to smooth the density contrast \cite{Ando:2018qdb}, the background equation of state when modes re-enter the horizon \cite{Musco:2012au} (the QCD transition can motivate a population of solar mass PBHs \cite{Byrnes:2018clq}) and the shape and sphericity of the initial energy-density profile \cite{Nakama:2013ica,Harada:2015yda,Kuhnel:2016exn,Musco:2018rwt}. More broadly, the general calculation has been questioned recently in \cite{Yoo:2018kvb,Germani:2018jgr}, with particular uncertainty on the critical density threshold $\delta_c$ and more than an order of magnitude uncertainty in  the relation between the horizon and PBH mass. There is not yet any consensus on how to calculate $\beta(M)$, given a particular primordial power spectrum.

Conditional on all of the aforementioned caveats, for a non-negligible number of PBHs to be generated, the amplitude of the power spectrum needs to be of order $10^{-2}$ depending on the mass of the PBH \cite{Cole:2017gle}. We will now see the relevant constraints.

\subsection{CMB Spectral distortions}

What we see in the CMB is a snapshot of acoustic excitations in the primordial plasma around the time of last scattering. Sound waves dissipate energy as they propagate, transferring energy into heating the ambient medium. Any heat dissipated into the primordial plasma has the possibility of showing up in the form of $\mu$ and $y$ type distortions of the CMB (\cite{Chluba:2012we, Chluba:2015bqa}), provided they occur in the redshift window between $z \sim 10^6$ and last scattering at $z \sim 10^3$. The reason for this is that at sufficiently early times, Compton scattering is efficient enough to rapidly restore thermal equilibrium after any energy injection process. At around $z \sim 10^6$ its efficiency starts to drop, and distortions of the blackbody spectrum of the CMB can start to persist if they were initially large enough. The greater the power spectrum is at small scales, the greater the amount of energy that gets dissipated into the primordial plasma, hence spectral distortions offer a powerful probe of the power spectrum at scales beyond those accessible in CMB anisotropies. Of the two varieties of distortions generated by dissipation, $\mu$-type distortions are sensitive to power at smaller scales. It corresponds to a distortion of the black body spectrum mimicked by an effective chemical potential $\mu$, given by \cite{Chluba:2013dna}
\eq{mu-dist}{\mu_\calR\approx  \int_{k_{\rm min}}^\infty \frac{d k}{k} \calP_\calR(k) \, W_{\calR,\mu}(k),
}
where the window function $W_\mu$ is given by
\eq{}{W_{\calR,\mu}(k) \approx  2.27 \left[
	\exp\left(-\left[\frac{\hat{k}}{1360}\right]^2\Bigg/\left[1+\left[\frac{\hat{k}}{260}\right]^{0.3}+\frac{\hat{k}}{340}\right]\right) 
	- \exp\left(-\left[\frac{\hat{k}}{32}\right]^2\right)
	\right],}
with $\hat{k}=k/[1\,{\rm Mpc}^{-1}]$ and $k_{\rm min}\simeq 1\,{\rm Mpc}^{-1}$, below which the power spectrum is tightly constrained by large scale observations of the CMB. 

Measurements from COBE/FIRAS require the $\mu$-distortion to be no greater than $9\times10^{-5}$ \cite{Fixsen:1996nj,Mather:1993ij}. For a reasonably broad peak with approximately $k^4$ slope, 
centred on $k\sim10^5{\rm Mpc^{-1}}$, the resulting $\mu$-distortion is $\mu\approx9\times10^{-7}$. This scale corresponds to larger black hole masses than those detected by LIGO and even then, the constraint is not under pressure. The largest possible PBH that can be produced and be consistent with the $\mu-$distortion constraint has mass $\sim4\times10^{4}M_\odot$ assuming that the PBH mass equals the horizon mass at the time of horizon entry. This is calculated assuming that the amplitude of the power spectrum is required to reach the current constraints from PBHs, shown by the orange line in Fig.~\ref{fig:new-bringmann}, which already rules out $f=\Omega_{\rm PBH}/\Omega_{\rm DM}=1$ on a large range of scales. See Fig.~\ref{fig:new-bringmann} for a plot of the full $\mu$-distortion constraints. Note that each point of the blue and purple lines represents the maximum allowed value of $A_s$ at the scale of the peak, $k_p$, for a delta function power spectrum $\mathcal{P}_\calR=A_s \delta(\log(k/k_p))$ (blue line) and a $k^4$ power spectrum $\mathcal{P}=4 A_s (k/k_p)^4$ cut off to zero for $k>k_p$ (purple line), after having integrated over each to find the total contribution to the $\mu$-distortion value.

We note that (\ref{mu-dist}) captures only the $\mu$-distortion induced by dissipation of scalar modes. Tensor modes can also produce dissipation distortions, with a resulting $\mu$-distortion given by $\mu_h\approx  \int_{k_{\rm min}}^\infty \frac{d k}{k} \calP_h(k) \, W_{h,\mu}(k)$,  \cite{Chluba:2014qia}. However the corresponding window function $W_{h,\mu}(k)$ is such that the overall distortion is some six orders of magnitude smaller for a nearly scale-invariant spectrum (independent of the amplitude), although it has much broader support, and is sensitive to power up to scales approaching $k \sim 10^5\, \mpc^{-1}$. As we review in the next subsection, scalar perturbations can source tensor perturbations at second order, and any enhancement of the primordial power spectrum at small scales will source enhanced tensor perturbations at commensurate scales. Although these offer no meaningful constraints with present day sensitivities, a PIXIE-like survey \cite{Kogut:2011xw} (with sensitivity to $\mu$-distortions as small as $\mu \sim 10^{-8}$) could be sensitive to primordial power spectrum enhancements of up to $\calP_\calR \sim 10^{-2}$ at $k \sim 10^{5} \mpc^{-1}$ due to the dissipation from secondarily produced tensors.

\subsection{Pulsar Timing Arrays}

\begin{figure}[H]
\centering
\includegraphics[width=10cm]{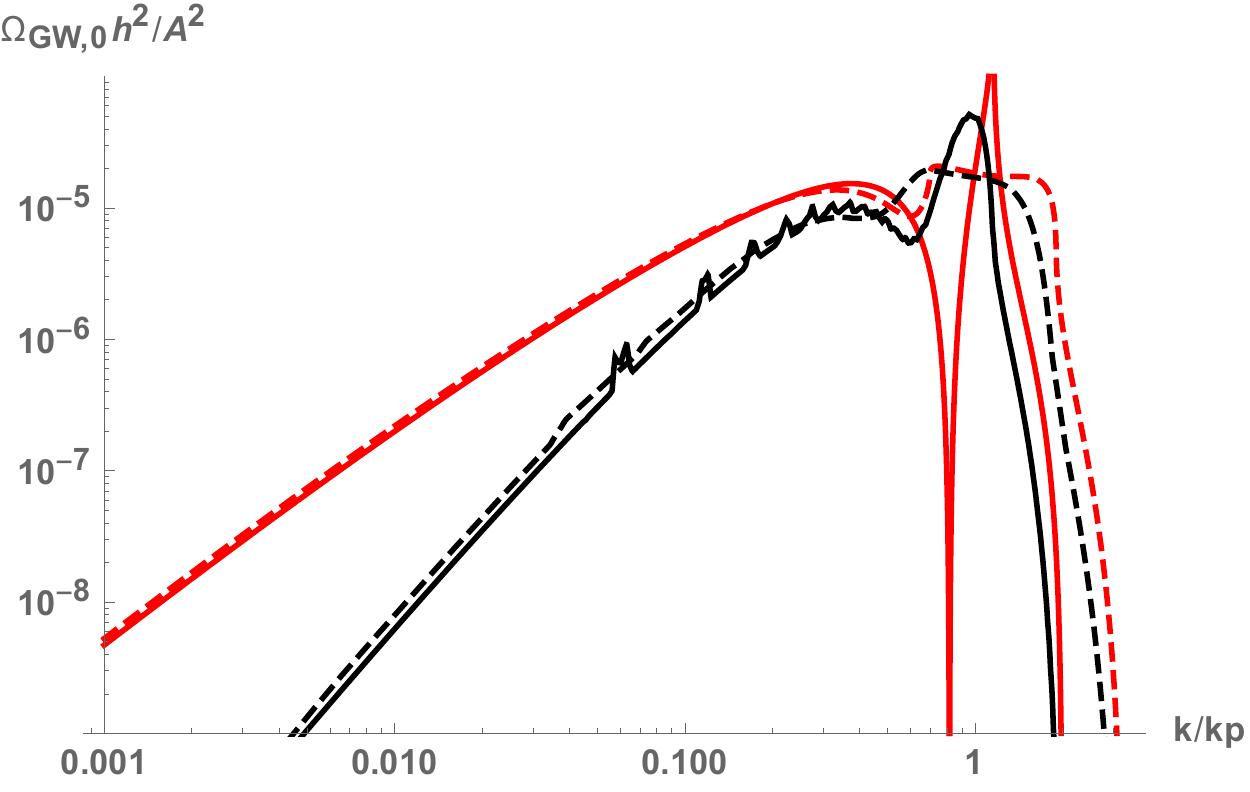}
\caption{The GW amplitude today for a delta function power spectrum and one growing like $k^4$ in red and black respectively, as described in the main text. The dashed lines are the values smoothed over 1 efolding. The small `teeth'-like features of the black line are due to numerical noise and they don't affect the power spectrum constraints derived from these curves.}
\label{OGW}
\end{figure}

Although scalar and tensor perturbations decouple at linear order, if the power spectrum is sufficiently boosted to generate PBHs a potentially observable amplitude of second-order gravitational waves (GWs) will be generated  \cite{Ananda:2006af,Osano:2006ew,Baumann:2007zm,Inomata:2016rbd,Garcia-Bellido:2017aan,Ando:2017veq,Saito:2008jc,Nakama:2016enz,Orlofsky:2016vbd,Gong,Caprini:2018mtu,Ando:2018nge,Kohri:2018awv}. This can be intuited as arising from interactions of the form $h_{ij}\partial_i\calR\partial_j\calR$. Specifically, the transverse traceless projection of the spatial part of the `stress-tensor'\footnote{By this, we simply mean the symmetric tensor obtained by varying the cubic interaction terms of the form $\calR\calR h$ in the perturbed action w.r.t. $h_{ij}$.} of the curvature perturbation can source tensor perturbations at second order, resulting in an induced contribution to the tensor power of the form
\eq{}{\calP_h(\tau,k) = \int_0^\infty dv\int_{|1-v|}^{1+v} du\, K(\tau, u,v)\calP_\calR(k u)\calP_\calR(k v)}
where $K(\tau,u,v)$ is a rapidly oscillating kernel whose precise form can be found in e.g. \cite{Inomata:2016rbd, Kohri:2018awv}. Given the convoluted nature of the integrand, many papers in the literature consider PTA constraints arising from a simple (though unphysical) delta function power spectrum but we also consider a more physical $k^4$ spectrum with a sharp cut-off.
In Fig.~\ref{OGW} we plot the GW amplitude for the two power spectra
\bea \calP_\calR(k)&=&A \delta(\log(k/k_p)), \\ 
\P_\calR(k)&=&  4A (k/k_p)^{4}  \;\;\;{\rm for}\;\;\; k<k_p ,\;\;\; 0\;\;\rm{otherwise}.  \eea
The factor of 4 is included in the latter power spectrum in order that both power spectra are normalised as $\int\P_\calR(k){\rm d}\ln k=A$. We also plot the GW spectrum averaged over an efolding (the dashed lines), because the gravitational wave energy of the delta function power spectrum diverges at $k=2k_p/\sqrt{3}$. The smoothed spectrum is defined by
\bea \Omega_{\rm GW}^{\rm smooth}(k)= \int^{ke^{1/2}}_{ke^{-1/2}}\Omega_{\rm GW}(k){\rm d}\ln k. \eea 
The amplitude of the smoothed spectra is similar near the peaks, and the key difference is the different scaling behaviour at small $k$. The delta function scalar power spectrum produces a gravitational wave spectrum which scales like $k^2$ while the  $k^4$ scalar power spectrum produces a gravitational wave spectrum which scales like $k^3$ at small $k$. This means that the constraints for values of $k_p$ larger than the scales which PTA best constrain will differ markedly for the two power spectra. The counterintuitive result that a delta function power spectrum does not give rise to the narrowest possible GW spectrum has been observed in numerous papers, e.g.~\cite{Bugaev:2009zh,Bartolo:2018rku}. Since the scalar power spectrum cannot grow faster than $k^4$, a $k^2$ tail in $\Omega_{\rm GW}$ cannot be produced by a narrowly peaked scalar power spectrum. It has been suggested that including non-Gaussianity in the calculation can mean that a delta-function or very narrow power spectrum will also induce a $k^3$ tail in $\Omega_{\rm GW}$ \cite{Cai:2018dig}. However, as we have shown that the power spectrum cannot grow that quickly, this observable would be indistinguishable from a $k^3$ tail produced by broader power spectra. We would however expect a turnover from the $k^3$ tail to a spectrum with broader support on small $k$ if the power spectrum is made significantly broader, since for a scale-invariant power spectrum one would expect a scale-invariant gravitational wave spectrum - this requires further investigation. Other effects of including non-Gaussianities are discussed in \cite{Unal:2018yaa}. To make the plot, we have used $\Omega_{\rm rad,0}h^2=4\times10^{-5}$ in order to evolve the GW amplitude from horizon entry during radiation domination until today.

There is, however, a discrepancy in the normalisation between various references in the literature that is most apparent when one tries to calculate the secondary tensor spectrum produced by a scale-invariant scalar power spectrum. In particular, the results of \cite{Ananda:2006af} (quoted in \cite{Caprini:2018mtu}), \cite{Baumann:2007zm} and \cite{Kohri:2018awv} differ, with the latter reference finding a normalisation that is order $10^{-2}$ less than the prior references. The precise source of this discrepancy is not immediately apparent to us. However, it is apparent that the numerical integrations necessary to arrive at the final answer are sufficiently involved as to make any analytic simplifications (such as those provided by \cite{Kohri:2018awv,Espinosa:2018eve}) advantageous. For this reason, we utilise the explicit analytic form of the kernel $K(\tau,u,v)$  detailed in \cite{Kohri:2018awv} in what follows -- for the purposes of placing observational bounds, it is also the more conservative choice because it leads to a lower normalisation of $\Omega_{\rm GW}$ than several other references. Using the simplifications provided by \cite{Kohri:2018awv, Espinosa:2018eve}\footnote{These references have resulting analytic forms for the kernel $K(\tau,u,v)$ that agree up to having taken different lower limits in eq.~(15) in \cite{Kohri:2018awv} and the corresponding eq.~(33) in \cite{Espinosa:2018eve}. The resulting difference will be negligible whenever the source scalar modes are sub-Hubble. We thank Davide Racco for correspondence on this matter.}, we can also evaluate the secondary tensor perturbations produced by a $k^4$ steepest growth spectrum.

For each frequency, we use the tightest constraint for a stochastic GW background from various PTA experiments \cite{Lentati:2015qwp,Arzoumanian:2015liz,Shannon:2015ect} in order to plot the power spectrum constraint in Fig.~\ref{fig:new-bringmann}. We use the unsmoothed GW spectrum induced by the  $k^4$  scalar power spectrum (in order to avoid needing to choose a smoothing scale which depends on the experiment) and convert from $k$ to frequency space using  $f=k c_s/(2 \pi)$ where $c_s=9.7\times 10^{-15}\Mpc/s$. Over the range of almost two orders of magnitude in $k$ (over 3 orders of magnitude in horizon mass from $\sim0.1M_\odot\,-\,200M_\odot$, see equation \eqref{ktoM}) the constraint on the power spectrum is stronger than the constraint from the non-detection of PBHs, meaning that PBHs will not form in significant numbers over this range of scales. We have discussed the many caveats at the beginning of this section. Notice that because the delta function power spectrum has a slower decay of $\Omega_{\rm GW}$ towards small $k$, the constraints on the power spectrum would become stronger than the $k^4$ spectrum constraints for scales sensitive to the low-frequency tail of $\Omega_{\rm GW}$.

\subsection{Implications for model building}\label{sec:model-building} 

\begin{figure}
\includegraphics[width=0.9\textwidth]{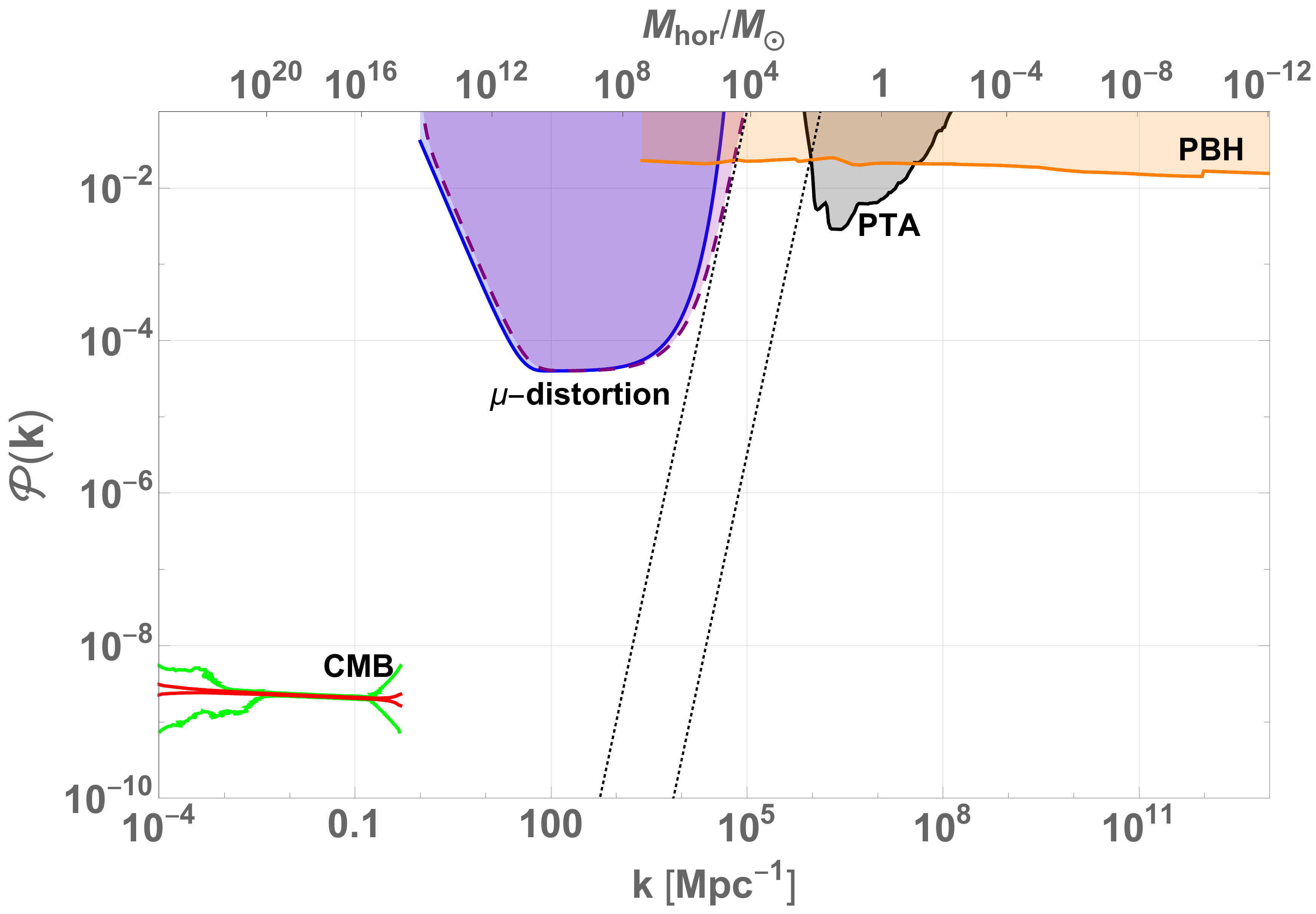}
\caption{Observational constraints on the power spectrum. The lines at small $k$ are the Planck 1$\sigma$ and $3\sigma$ measurements. On much smaller scales there are only upper bounds; shaded regions are disallowed. The solid blue line shows the upper bound from $\mu$-distortions for a delta function power spectrum, $\mathcal{P}_\calR=A_s \delta(\log(k/k_p))$, as a function of $k_p$, and the solid orange line shows the PBH upper bounds, subject to the uncertainties discussed in the main text. The dashed purple line shows the upper bound from $\mu$-distortions for the steepest growth power spectrum $\mathcal{P}_\calR=4 A_s (k/k_p)^4$ which drops to zero for $k>k_p$, and the solid black line shows the PTA upper bounds for the same power spectrum. The factor of $4$ is included so that it has the same normalisation as the delta function power spectrum when integrating with respect to $\ln k$. The dashed black lines have a $k^4$ slope.}
\label{fig:new-bringmann}
\end{figure}

In Fig.~\ref{fig:new-bringmann} we produce a new plot of the constraints on the primordial power spectrum on all scales similar to Bringmann et al.~\cite{Bringmann:2011ut}, but unlike \cite{Bringmann:2011ut} we do not include constraints from ultracompact minihalos (UCMHs) since they rely on a WIMP DM scenario. For a discussion of recent UCMH constraints see \cite{Gosenca:2017ybi,Delos:2017thv,Nakama:2017qac,Delos:2018ueo}, and mixed scenarios with both WIMPs and PBHs are discussed in \cite{Lacki:2010zf,Eroshenko:2016yve,Boucenna:2017ghj}. 

On the largest scales we plot the Planck measurements of the power spectrum \cite{Akrami:2018vks}. The next relevant constraints on smaller scales come from $\mu$-distortions. Since the power spectrum cannot grow arbitrarily quickly, it is clear that the power spectrum cannot become large enough to generate PBHs on scales $k<  10^4 \Mpc^{-1}$, subject to the aforementioned assumptions, the most relevant being that the perturbations are Gaussian \cite{Nakama:2017xvq}.  Hence there is no need to also show the $y$-distortion constraints which affect larger scales. The blue line is the upper bound on the amplitude for a monochromatic power spectrum, whilst the dashed purple line is the upper bound on the amplitude for a power spectrum with $k^4$ slope and immediate drop off. For a constraint on slightly smaller scales than spectral distortions, see \cite{Jeong:2014gna}.

The black line on scales $k\sim10^7\Mpc^{-1}$ represents the upper bound due to the PTA constraints, while the relatively flat orange line represents the PBH constraint. The PBH constraints are calculated using values of $f=\Omega_{\rm PBH}/\Omega_{\rm DM}$ from \cite{Inomata:2017vxo} and \cite{Carr:2017edp} for PBH masses between $\sim10^{-24}{\rm M}_\odot$ and $10^{7}{\rm M}_\odot$. These combine various constraints from e.g.~their evaporation, femto-lensing of gamma-ray bursts, neutron-star capture, white dwarf explosions, and microlensing. We use $\delta_c=0.45$ for definiteness, and we effectively use a delta function for the window function in translating the variance of the perturbations to the amplitude of the power spectrum. We also use a one-to-one relation between the density contrast $\delta$ and the comoving curvature perturbation $\mathcal{R}$, which is not realistic. We do not include constraints from microlensing for masses $\lesssim10^{-10}M_\odot$ due to uncertainties concerning the wave effect \cite{Inomata:2017vxo}. The slight dip at $k\sim10^7$ is caused by including the effect of the change in the equation of state around the time of the QCD transition \cite{Byrnes:2018clq}. Also note that we use
\eq{ktoM}{\frac{k}{3\times10^{22}\,{\rm Mpc^{-1}}}=\frac{\pi}{t_ic}\left(\frac{t_i}{t_{\rm eq}}\right)^\frac{1}{2}\left(\frac{t_{\rm eq}}{t_0}\right)^\frac{2}{3}\left(\frac{g_{*,i}}{g_{*,0}}\right)^{-\frac{1}{12}},\qquad{t_i=\frac{10^{-38}(M/1{\rm g})}{\gamma}{\rm s}}}
to convert between $k$ and PBH mass, where $t_i$ is horizon entry time of the overdensity, $M$ is the PBH mass, $\gamma$ is the fraction of the horizon mass that will collapse to form the black hole which we take to be 1 given the uncertainty in the literature, $t_{\rm eq}\approx2\times10^{12}\,{\rm s}$ is the cosmic time of matter-radiation equality and $t_0\approx4\times10^{17}\,{\rm s}$ is the cosmic time today. We take the effective degrees of freedom today to be $g_{*,0}\approx3.36$ and $g_{*,i}$ is the effective degrees of freedom at the time of horizon entry.

In order to reach the current constraint on the number of PBHs (orange line) from the amplitude of the power spectrum at CMB scales, the growth must begin at $k\gtrsim10^3\,{\rm Mpc^{-1}}$ in order to avoid constraints from the $\mu$-distortions, since it can only grow as fast as $k^4$. This is shown by the left-most dotted black line in Fig.~\ref{fig:new-bringmann}. This implies a maximum PBH mass which can be generated of $4\times10^4M_\odot$ corresponding to $k\sim7\times10^4\,{\rm Mpc^{-1}}$. This point is where the left dashed black line (with $k^4$ slope) crosses the PBH constraint line, and it is also where the dashed purple line which marks the distortion constraints for a $k^4$ growth crosses the PBH constraint line. Notice that the blue line (for a delta function power spectrum) crosses the PBH constraint line at a larger scale $k\sim4\times10^4\,{\rm Mpc^{-1}}$, corresponding to a PBH mass of $2\times10^5\,M_\odot$. The difference between these two masses demonstrates the additional restriction on PBHs caused by the restriction on the steepest possible growth of the power spectrum. 

Similarly, in order to avoid PTA constraints, the power spectrum growth must begin at $k\lesssim10^4\,{\rm Mpc^{-1}}$ as shown by the right-most dotted black line in Fig.~\ref{fig:new-bringmann}. This assumes that the power spectrum can drop off instantaneously to $\mathcal{P}_\mathcal{R}\lesssim10^{-3}$ after the peak to avoid the PTA constraints -- see App.~\ref{sec:end} for a discussion on this point. For PBHs with masses larger than those constrained by PTA experiments, the power spectrum is free to grow from $k\gtrsim6\times10^5\,{\rm Mpc^{-1}}$, as a $k^4$ slope will clear the PTA constraints from this scale, and there are no severe constraints on the power spectrum on smaller scales. In order to produce PBHs on a scale $k\sim10^6\,\mpc^{-1}$ and avoid the $\mu$-distortion constraint, the power spectrum needs to grow at least as steeply as $k^{1.2}$ on the scales between those two constraints.

Early matter-dominated scenarios are of interest because the lack of pressure means that PBHs are able to form much more easily and have been considered recently in e.g.~\cite{Harada:2016mhb,Carr:2017edp,Cole:2017gle,Georg:2016yxa}. This means that the amplitude of the power spectrum is related to the number of PBHs by a power law instead of logarithmically as is the case in radiation domination that we have assumed to plot the orange line in Fig.~\ref{fig:new-bringmann}. One could then question whether the constraints on the power spectrum change more quickly than the $k^4$ limit. Using constraints on the power spectrum from \cite{Cole:2017gle}, we have verified that they do not change more quickly than $k^4$, and therefore that PBHs of every possible mass can still be generated while respecting this bound on the power spectrum growth. 

\section{Conclusions}\label{conc}

We have shown that the steepest possible growth of the primordial power spectrum is given by $n_s-1=4$ during canonical single-field inflation, independent of the shape of the inflaton potential. Such a rapid growth is only possible when the inflaton makes a rapid transition from ``slow-roll'' inflation to non-attractor inflation, characterised by an almost exactly flat potential, and remains true even if the potential is not always decreasing\footnote{Note added: As we were preparing this paper, ref.~\cite{Cheng:2018qof} appeared, aiming to derive a lower bound of $\eta > -6$ from causality arguments. However, the matching calculation on which this is based neglected to impose both Israel junction conditions (\ref{jc}), imposing only the continuity of $\calR_k$. Furthermore, the correct causality criteria one should impose is that the commutator of the curvature perturbation at two points should vanish at space-like separation, trivially satisfied even when matching with an intermediate phase of $\eta < -6$.}. In the standard case of single-clock inflation - implying the curvature perturbation freezes out shortly after horizon crossing - the power spectrum grows less steeply, and is bounded by $n_s-1<3$. It would be of great interest to understand whether our bound can be violated in more complicated models of inflation. 
For example, see \cite{Ando:2018qdb}, which in some cases requires an ad hoc power spectrum with steepness up to $k^8$ in order to evade power spectrum constraints while generating PBHs, which our bound implies is not possible in the context of single-field inflation. 

We have calculated analytic expressions for the most rapid growth of the power spectrum possible, by matching the curvature perturbation between various phases of inflation, characterised by different rates at which the slow-roll parameter $\epsilon$ decreases. The steep $k^4$ growth arises during times when modes exiting the horizon are affected by both periods of inflation. We have also provided a way to reconstruct the inflaton potential given an arbitrary time evolution of the expansion rate during inflation specified by $\epsilon(t)$ in App.~\ref{appendix:reconst}.

Due to the phenomena of critical collapse to form PBHs, the PBH mass spectrum cannot be arbitrarily close to monochromatic. We have shown that the mass spectrum is remarkably insensitive to the shape of the power spectrum close to its peak amplitude, with everything from a gentle growth, $n_s-1=0.1$ to the extreme (and impossible) case of a delta function power spectrum producing a comparable width for the mass function of PBHs. 

In Fig.~\ref{fig:new-bringmann} we have combined the key measurements and constraints on the primordial power spectrum, showing how on various scales, CMB measurements, CMB spectral distortion constraints or PTA constraints all force the power spectrum to be too small to generate PBHs. There does however remain a window between the latter two constraints which is sufficiently broad such that the power spectrum can grow and produce large PBHs without conflicting either of those constraints and without requiring the perturbations to be non-Gaussian. 

\begin{figure}
\includegraphics[width=\textwidth]{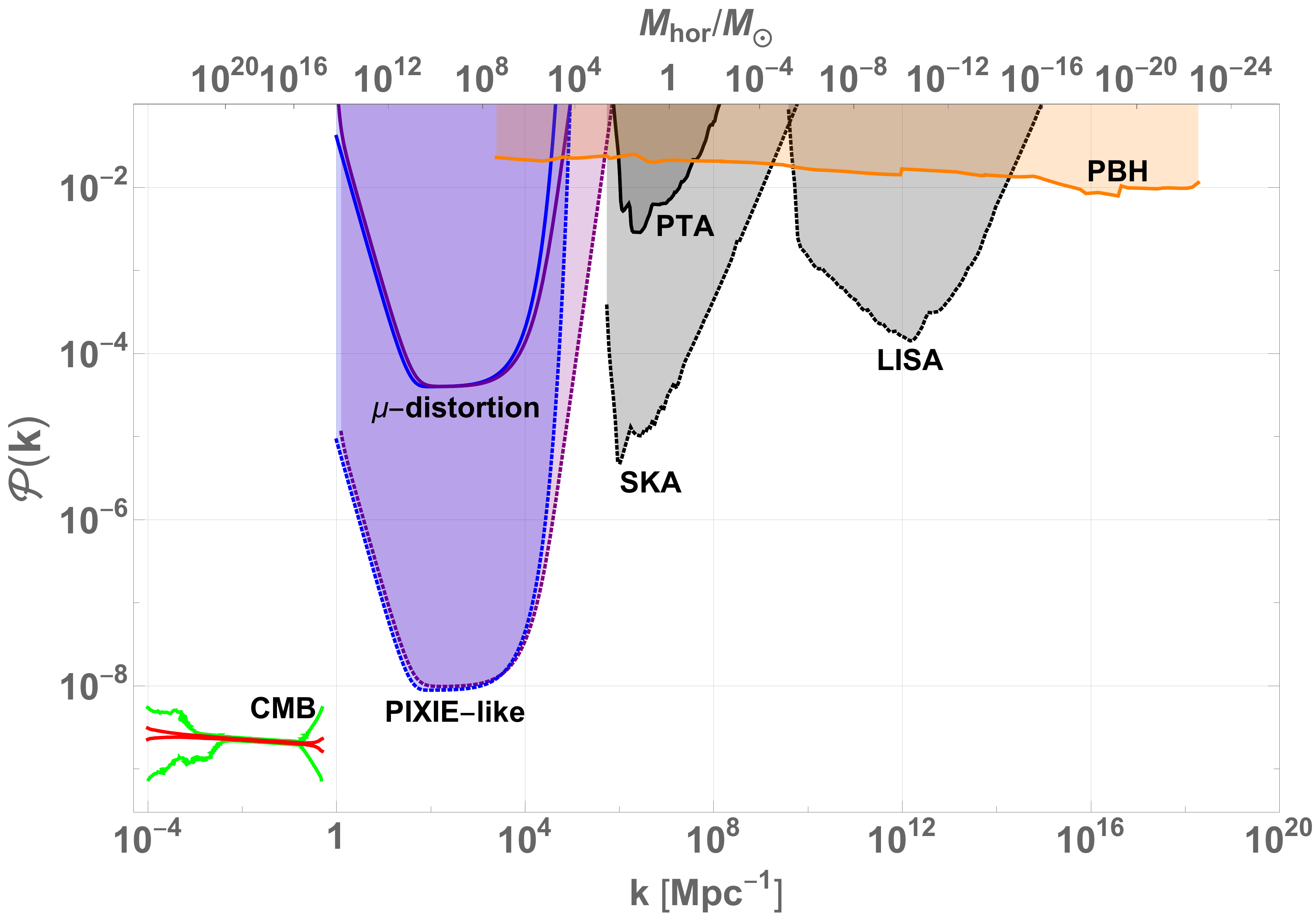}
\caption{Current and forecasted constraints on the amplitude of the power spectrum. The solid lines are the same as in Fig. \ref{fig:new-bringmann}, apart from the x-axis which is extended to the smallest scales that PBHs constrain, corresponding to the horizon scale which generates a PBH that decays during big bang nucleosynthesis. The dashed lines show forecasted future constraints from a PIXIE-like satellite for $\mu$-distortions (the dashed blue line assumes a delta function power spectrum while the purple line has a power spectrum growing at the maximum rate of $k^4$), and the dashed black lines are induced gravitational wave forecasts for a $k^4$ scalar power spectrum with a cut off, using PTA constraints from SKA and from the LISA satellite on smaller scales. Shaded regions are disallowed.}
\label{fig:forecast}
\end{figure}

We plot forecasted constraints on the power spectrum in Fig.~\ref{fig:forecast}. The sensitivity curves for SKA and LISA are extracted from \cite{Moore:2014lga} and do not include the possible degradation due to astrophysically generated gravitational waves. LISA covers the scales corresponding to a possible window where PBHs could consist of all the DM, with masses in the range $M_{\rm pbh}\sim10^{-13}-10^{-7}M_\odot$  \cite{Saito:2008jc}.  Of particular interest is how the gap between future $\mu$-distortion constraints assuming a PIXIE-like experiment which can probe $\mu=2\times10^{-8}$ and the existing PTA constraints becomes a factor of $2$ in $k$-space, corresponding to less than an e-folding of inflation, which in practice means constructing a model which grows at the maximum rate and then decreases again is unrealistic. A more detailed treatment of the PTA constraint at low frequency would probably completely close the gap, and the addition of SKA constraints from pulsars does close the gap. The difference of a factor of 8 in $k$ between where the two different forecasted $\mu$-distortion lines cross the PBH line show how much more powerful the PIXIE constraint on PBHs becomes once including the maximum growth rate of the power spectrum. Therefore, PIXIE combined with PTA constraints and the steepest growth rate that we have derived would be able to rule out the generation of LIGO mass PBHs, unless the initial perturbations are sufficiently non-Gaussian on the relevant range of scales. Finally, the combined constraints from the CMB, a PIXIE-like experiment, SKA and LISA will almost completely rule out Gaussian perturbations being able to generate any PBHs with masses greater than $10^{-15}M_\odot$.

\textbf{Note added:} after submitting the first version of our paper, \cite{Inomata:2018epa} appeared which deals with current and future constraints on induced gravitational waves. We would like to thank the authors for helpful discussions and comments on both of our papers.

\section{Acknowledgements}

We thank David Seery for extensive help with his CppTransportCode \cite{Seery:2016lko}. CB is a Royal Society university research fellow. SP is supported by funds from Danmarks Grundforskningsfond under grant no.~1041811001. PC acknowledges support from the UK Science and Technology Facilities Council via Research Training Grant ST/N504452/1. We are grateful to Cliff Burgess, Jens Chluba, Dani Figueroa, Kazunori Kohri, Ilia Musco, Davide Racco, David Seery, Tommi Tenkanen, Takahiro Terada, Jennie Traschen, Michael Trott, and Sam Young for valuable discussions over the course of this investigation. We acknowledge the Centro de Ciencias de Benasque Pedro Pascual for providing an excellent working environment during the 2017 Understanding Cosmological Observations meeting where this work was initiated.
 
\appendix

\section{The primordial power spectrum from matching}\label{appendix:matching}

It is possible to arrive at an analytic understanding of various features of the shape of the primordial power spectrum generated by transiting into and out of a phase of ultra slow-roll (USR) inflation by approximating the evolution of $\eta$ as a series of phases of constant $\eta$, and matching between these phases. In this appendix, we'll perform a series of matchings, culminating in a four-stage matching from $\eta \equiv 0 \to -2 \to -6 \to 2 \to 0$.  

We begin by matching from $\eta=0$ to $\eta=-6$ (USR) and back to $\eta=0$. For this we need the mode functions for inflation for each phase, which are obtained from the equation of motion for the Mukhanov-Sasaki variable, $\upsilon_k=z\mathcal{R}$, where $z^2 = 2a^2\mpl^2\epsilon$ and $\mathcal{R}$ is the comoving curvature perturbation:
\eq{eq:MS}{\upsilon_k^{\prime\prime}+(k^2-\frac{z^{\prime\prime}}{z})\upsilon_k=0.}
By directly differentiating $z$ with respect to conformal time, it can be shown that 
\eq{eq:zprime}{\frac{z^{\prime\prime}}{z}=(aH)^2\left(2-\epsilon+\frac{3}{2}\eta+\frac{1}{4}\eta^2-\frac{1}{2}\epsilon\eta+\frac{1}{2}\frac{\dot{\eta}}{H}\right).}
Equation (\ref{eq:zprime}) is exact to all orders. Assuming that $\epsilon\ll1$, we can rewrite equation (\ref{eq:MS}) as 

\eq{eq:ups}{\upsilon_k^{\prime\prime}+(k^2-\frac{\nu^2-\frac{1}{4}}{\tau^2})\upsilon_k=0,}
with a new parameter $\nu$ defined as
\eq{eq:nu}{\nu^2=\frac{9}{4}+\frac{3}{2}\eta+\frac{1}{4}\eta^2+\frac{\dot{\eta}}{2H}.}
The solutions for the canonically normalised mode function are of the form
\eq{msv}{v_k = \frac{\sqrt{\pi}}{2}e^{i\left(\nu + \frac{1}{2}\right)\frac{\pi}{2}}\sqrt{-\tau}H^{(1)}_\nu(-k\tau)  }
in linear combination with its complex conjugate, where $H^{(1)}_\nu(-k\tau)$ is the Hankel function of the first kind. 

For a constant $\eta$ phase, the last term in (\ref{eq:nu}) vanishes, and we find that for $\eta = -6$ and $\eta = 0$, $\nu^2 = 9/4$, with $\nu = -3/2$ corresponding to USR and $\nu = 3/2$ corresponding to SR. The curvature perturbation is obtained via $\calR_k = v/z$ with $z^2 = 2a^2\mpl^2\epsilon$, and we now find the mode equations for each phase. For $\nu = 3/2$ (phase 1, $\eta=0$), the curvature perturbation is given by
\eq{SR}{\calR^{(1)}_k = i\frac{H}{\mpl}\frac{1}{\sqrt{4\epsilon_1 k^3}}\left[c (1 + ik\tau)e^{-ik\tau} - s (1 - ik\tau)e^{ik\tau} \right],}
where $\epsilon(\tau)=\epsilon_1$ is treated as constant and $c,s$ are constant coefficients to be found via the matching, and they should satisfy the Wronskian condition. This needs to be matched to the Bunch-Davies vacuum in the limit $\tau\rightarrow-\infty$, so the mode equation during the first phase of $\eta=0$ reduces to 
\eq{eta0mode}{\mathcal{R}^{(1)}_k=i\frac{H}{\mpl}\frac{e^{-ik\tau}}{\sqrt{4\epsilon_1k^3}}(1+ik\tau)}
i.e.~$c=1$ and $s=0$. Using the relation $H^{(1)}_{-3/2} = -i H^{(1)}_{3/2}$, and writing $\epsilon(\tau)$ during USR as 
\eq{}{\epsilon(\tau) = \epsilon_1\left(\frac{a(\tau_1)}{a(\tau)}\right)^6=\epsilon_1\left(\frac{\tau}{\tau_1}\right)^6}
where $\tau_1$ is the time of transition between $\eta=0$ and $\eta=-6$ and the second equality comes from $aH=-1/\tau$, we find the canonically normalised mode functions during the phase of $\eta=-6$ (phase 2, USR) to be
\eq{USR}{\calR^{(2)}_k = i\frac{H}{\mpl}\frac{(\tau_1/\tau)^3}{\sqrt{4\epsilon_1 k^3}}\left[c_1 (1 + ik\tau)e^{-ik\tau} - s_1 (1 - ik\tau)e^{ik\tau} \right],}
where $\tau_1$ and $\epsilon_1$ are fixed, and the coefficients $c_1$ and $s_1$ will be determined by the matching from the $\eta=0$ phase, and hence will be in terms of $k$ and $\tau_1$. 

The matching conditions between the two phases are given by the Israel junction conditions \cite{Israel:1966rt, Deruelle:1995kd}
\eq{jc}{\left[\calR_k \right]_\pm = 0,\; \left[z^2 \calR'_k \right]_\pm = 0.}
The first of these is determined by requiring that the metric be continuous across the transition. The second follows from the equation for the mode function $\calR_k$:
\eq{}{\calR_k'' + \frac{(z^2)'}{z^2}\calR_k' + k^2\calR_k = 0,}
which implies that
\eq{}{\left(z^2\calR_k'\right)' = -z^2 k^2\calR_k.}
Integrating the above over an infinitesimal interval around the transition and recalling the continuity of $\calR_k$ and $z^2$ results in the second condition in (\ref{jc}). So, continuity between $\calR^{(1)}_k$ and $\calR^{(2)}_k$ at $\tau_1$ results in the equation
\eq{}{(1 + ik\tau_1)e^{-ik\tau_1} =c_1 (1 + ik\tau_1)e^{-ik\tau_1} - s_1(1 - ik\tau_1)e^{ik\tau_1}, }
and continuity of the time derivatives $\calR^{\prime(1)}_k$ and $\calR^{\prime(2)}_k$ at $\tau_1$ requires
\begin{eqnarray}
k^2\tau_1 e^{-ik\tau_1} = c_1 e^{-ik\tau_1}\left(k^2\tau_1 - \frac{3}{\tau_1}\left(1 + ik\tau_1\right)\right)  - s_1 e^{ik\tau_1}\left(k^2\tau_1 - \frac{3}{\tau_1}\left(1 - ik\tau_1\right)\right),
\end{eqnarray}
which together imply that
\eq{s}{s_1 =  \frac{3\,i\,e^{-2ik\tau_1}}{2(k\tau_1)^3}(1 + ik\tau_1)^2   }
and
\eq{c}{c_1 =  1+ \frac{3i(1 + k^2\tau_1^2)}{2(k\tau_1)^3}   .}
We see that in the limit $\tau_1 k \to -\infty$, that is for modes that are deep within the Hubble radius at $\tau_1$, $\theta_k \to 0$, which means that the corresponding modes are still in the BD vacuum during USR.

In order to meaningfully talk about a late time power spectrum, we need to end USR, otherwise the modes grow unboundedly. To model this, we consider a transition from USR back to a phase of $\eta=0$. In the final phase, the mode function corresponds to the usual case (\ref{SR}) but with constant $\epsilon$ given by $\epsilon_2 = \epsilon_1 (a_1/a_2)^6=\epsilon_1 (\tau_2/\tau_1)^6$, where $\log(a_2/a_1)$ is the total number of e-foldings of USR:
\eq{endphase}{\calR^{(3)}_k = i\frac{H}{\mpl}\frac{(\tau_1/\tau_2)^3}{\sqrt{4\epsilon_1 k^3}}\left[c_2 (1 + ik\tau)e^{-ik\tau} - s_2 (1 - ik\tau)e^{ik\tau} \right].}
We therefore need to compute another matching between the mode functions in (\ref{USR}) and (\ref{endphase}) at time $\tau_2$, which is when USR ends, in order to determine $c_2$ and $s_2$ which will be functions of $k$ and $\tau_2$. From requiring continuity of $\calR_k$ at $\tau_2$, we find that
\eq{}{c_2 - s_2 \frac{1 - i k \tau_2}{1 + i k \tau_2}e^{2ik\tau_2} =  c_1 - s_1 \frac{1 - i k \tau_2}{1 + i k \tau_2}e^{2ik\tau_2}}
whereas continuity of $\calR_k'$ at $\tau_2$ implies
\begin{eqnarray}
c_2 k^2\tau_2\, - s_2 k^2\tau_2\, e^{2ik\tau_2} &=&c_1 \left[k^2\tau_2 - \frac{3}{\tau_2}(1 + ik\tau_2)\right]\\ \nn &-& s_1 e^{2ik\tau_2}\left[k^2\tau_2 - \frac{3}{\tau_2}(1 - ik\tau_2)\right], 
\end{eqnarray}
which gives
\begin{eqnarray}\nn
c_2 &=&
-\frac{1}{4 k^6 \tau _1^3 \tau _2^3} \bigl\{9 e^{2 i k (\tau _2-\tau_1)} \left(k \tau _1-i\right){}^2 \left(k \tau _2+i\right){}^2- \left(k^2 \tau _1^2 \left(2 k \tau _1+3 i\right)+3 i\right) \left(k^2 \tau _2^2 \left(2 k \tau _2-3 i\right)-3 i\right)\bigr\}\\
\end{eqnarray}
and
\begin{eqnarray}\nn
s_2 &=&\frac{ e^{-2 i k \left(\tau _1+\tau _2\right)}}{4 k^6 \tau _1^3 \tau _2^3} \bigl\{3 e^{2 i k \tau _2} \left(3+k^2 \tau _2^2 \left(3-2 i k \tau _2\right)\right) \left(k \tau _1-i\right){}^2+3 i e^{2 i k \tau _1} \left(k^2 \tau _1^2 \left(2 k \tau _1+3 i\right)+3 i\right) \left(k \tau _2-i\right){}^2\bigr\}\\.
\end{eqnarray}

The power spectrum for the curvature perturbation at late times (during slow roll again) is 
\eq{srusr}{\calP_\calR = \lim_{\tau \to 0^-}\frac{k^3}{2\pi^2}|\calR_k^{(3)}|^2 =  \frac{H^2}{8\pi^2\mpl^2\epsilon_3}\left[c_2^*c_2 + s_2^*s_2 - s_2^*c_2 - s_2c_2^*\right],}
where 
\eq{}{\epsilon_3 := \epsilon_1\,e^{-6N_{\rm USR}}}
is the fixed, final value of $\epsilon$ during the second phase of $\eta=0$, determined by $\epsilon_1$ during the initial phase of $\eta=0$, and $N_{\rm USR}$ which is the total number of e-folds of USR. The resulting late time spectrum is the black line plotted in Fig.~\ref{fig:different-etas} with $N_{USR}=2.3$. 

We generalise this matching to go from $\eta=0$ to arbitrary constant $\eta<0$ and back to $\eta=0$ in order to plot the other lines in Fig.~\ref{fig:different-etas}. We do this in the same way as just described for $\eta=0$ to $\eta=-6$ and back to $\eta=0$, but replace the mode equation $\calR_k^{(2)}$ with the appropriate solution from equation (\ref{msv}) for each value of $\nu$ using 
\eq{}{\nu^2 = \frac{9}{4} + \frac{3\eta}{2} + \frac{\eta^2}{4} = \left(\frac{3 + \eta}{2}\right)^2}
for constant $\eta$. We also note that 
\eq{}{\epsilon  = \epsilon_1\left(\frac{a}{a_1}\right)^\eta=\epsilon_1\left(\frac{\tau_1}{\tau}\right)^\eta}
during the constant $\eta$ phase. We then do the matching using exactly the same method as before, and find that the late time power spectra plotted in Fig.~\ref{fig:different-etas} are finally given by:
\eq{srusr}{\calP_\calR = \lim_{\tau \to 0^-}\frac{k^3}{2\pi^2}|\calR_k^{(3)}|^2 =  \frac{H^2}{8\pi^2\mpl^2\epsilon_3}\left[c_2^*c_2 + s_2^*s_2- s_2^*c_2 - s_2c_2^*\right],}
with $\epsilon_3=\epsilon_1e^{-\eta{N_{\eta=const}}}$, where $N_{\eta=const}$ is set by the duration of the constant $\eta$ phase and the coefficients $c_2$ and $s_2$ are given in general by:

\small\begin{equation}
\label{c2}
c_2=\frac{i \pi  e^{-i k (\tau_1-\tau_2)}}{8 k^\frac{13}{2} \sqrt{-\tau_1} \sqrt{-\tau_2}} \left(\begin{split} & \sqrt{k^5} (k \tau_2+i) \left(\begin{split} & \left(k^4 \tau_1 H_{\frac{\eta+3}{2}}^{(1)}(-k \tau_1)+k^3 (1+i k \tau_1) H_{\frac{\eta+1}{2}}^{(1)}(-k \tau_1)\right) H_{\frac{\eta+1}{2}}^{(2)}(-k
   \tau_2) \\
   & +\left(-k^4 \tau_1 H_{\frac{\eta+3}{2}}^{(2)}(-k \tau_1)+k^3 (-1-i k \tau_1) H_{\frac{\eta+1}{2}}^{(2)}(-k \tau_1)\right) H_{\frac{\eta+1}{2}}^{(1)}(-k \tau_2) \end{split} \right) \\ 
   & +k^\frac{7}{2} \tau_2 \left(\begin{split} & \left(k^3 (k
   \tau_1-i) H_{\frac{\eta+1}{2}}^{(2)}(-k \tau_1)-i k^4 \tau_1 H_{\frac{\eta+3}{2}}^{(2)}(-k \tau_1)\right) H_{\frac{\eta+3}{2}}^{(1)}(-k \tau_2)\\
   & +\left(k^3 (-k \tau_1+i) H_{\frac{\eta+1}{2}}^{(1)}(-k \tau_1)+i k^4
   \tau_1 H_{\frac{\eta+3}{2}}^{(1)}(-k \tau_1)\right) H_{\frac{\eta+3}{2}}^{(2)}(-k \tau_2)\end{split}\right)\end{split}\right)
   \end{equation}

\begin{equation}
\label{s2}
s_2=-\frac{\pi  e^{-i k (\tau_1+\tau_2)}}{8 k^\frac{7}{2}
   \sqrt{-\tau_1} \sqrt{-\tau_2}} \left(\begin{split}& i k^\frac{7}{2} \tau_1 H_{\frac{\eta+1}{2}}^{(1)}(-k \tau_1) H_{\frac{\eta+1}{2}}^{(2)}(-k \tau_2)+i k^\frac{7}{2} \tau_2 H_{\frac{\eta+1}{2}}^{(1)}(-k \tau_1) H_{\frac{\eta+1}{2}}^{(2)}(-k
   \tau_2)\\
   & +\sqrt{k^5} H_{\frac{\eta+1}{2}}^{(1)}(-k \tau_1) H_{\frac{\eta+1}{2}}^{(2)}(-k \tau_2)+i k^{\frac{9}{2}} \tau_1 \tau_2 H_{\frac{\eta+3}{2}}^{(1)}(-k \tau_1) H_{\frac{\eta+1}{2}}^{(2)}(-k \tau_2)\\
   & +i k^\frac{9}{2} \tau_1
   \tau_2 H_{\frac{\eta+1}{2}}^{(1)}(-k \tau_1) H_{\frac{\eta+3}{2}}^{(2)}(-k \tau_2)-\sqrt{k^5} k^2 \tau_1 \tau_2 H_{\frac{\eta+1}{2}}^{(1)}(-k \tau_1) H_{\frac{\eta+1}{2}}^{(2)}(-k \tau_2)\\
   & +k^\frac{9}{2} \tau_1 \tau_2
   H_{\frac{\eta+3}{2}}^{(1)}(-k \tau_1) H_{\frac{\eta+3}{2}}^{(2)}(-k \tau_2)+\sqrt{k^3} k^2 \tau_1 H_{\frac{\eta+3}{2}}^{(1)}(-k \tau_1) H_{\frac{\eta+1}{2}}^{(2)}(-k \tau_2)\\
   & +k^\frac{7}{2} \tau_2 H_{\frac{\eta+1}{2}}^{(1)}(-k \tau_1)
   H_{\frac{\eta+3}{2}}^{(2)}(-k \tau_2)\\
   & +k^2 \tau_2 \left(-\sqrt{k^5} \tau_1 H_{\frac{\eta+3}{2}}^{(2)}(-k \tau_1)+\sqrt{k^3} (-1-i k \tau_1) H_{\frac{\eta+1}{2}}^{(2)}(-k \tau_1)\right) H_{\frac{\eta+3}{2}}^{(1)}(-k
   \tau_2)\\
   & +H_{\frac{\eta+1}{2}}^{(1)}(-k \tau_2) \left(\sqrt{k^5} (k \tau_1-i) (k \tau_2-i) H_{\frac{\eta+1}{2}}^{(2)}(-k \tau_1)-k^\frac{7}{2} \tau_1 (1+i k \tau_2) H_{\frac{\eta+3}{2}}^{(2)}(-k \tau_1)\right)\end{split}\right).
   \end{equation}
\normalsize   
We now move on to a more realistic matching, wherein one transitions in and out of USR with intermediate phases that interpolate between USR and SR.

\subsection{SR $\to \eta \equiv -2 \to$ USR $\to \eta \equiv 2 \to$ SR matching}\label{sect:-2}
By now, we see that the matching calculations involve nothing but sequentially solving a series of linear equations. We now attempt to model two additional intermediate phases to transition into USR, via a phase of $\eta = -2$, and out of USR, via a phase of $\eta=2$. 

\begin{figure}
\centering
\includegraphics[width=0.9\textwidth]{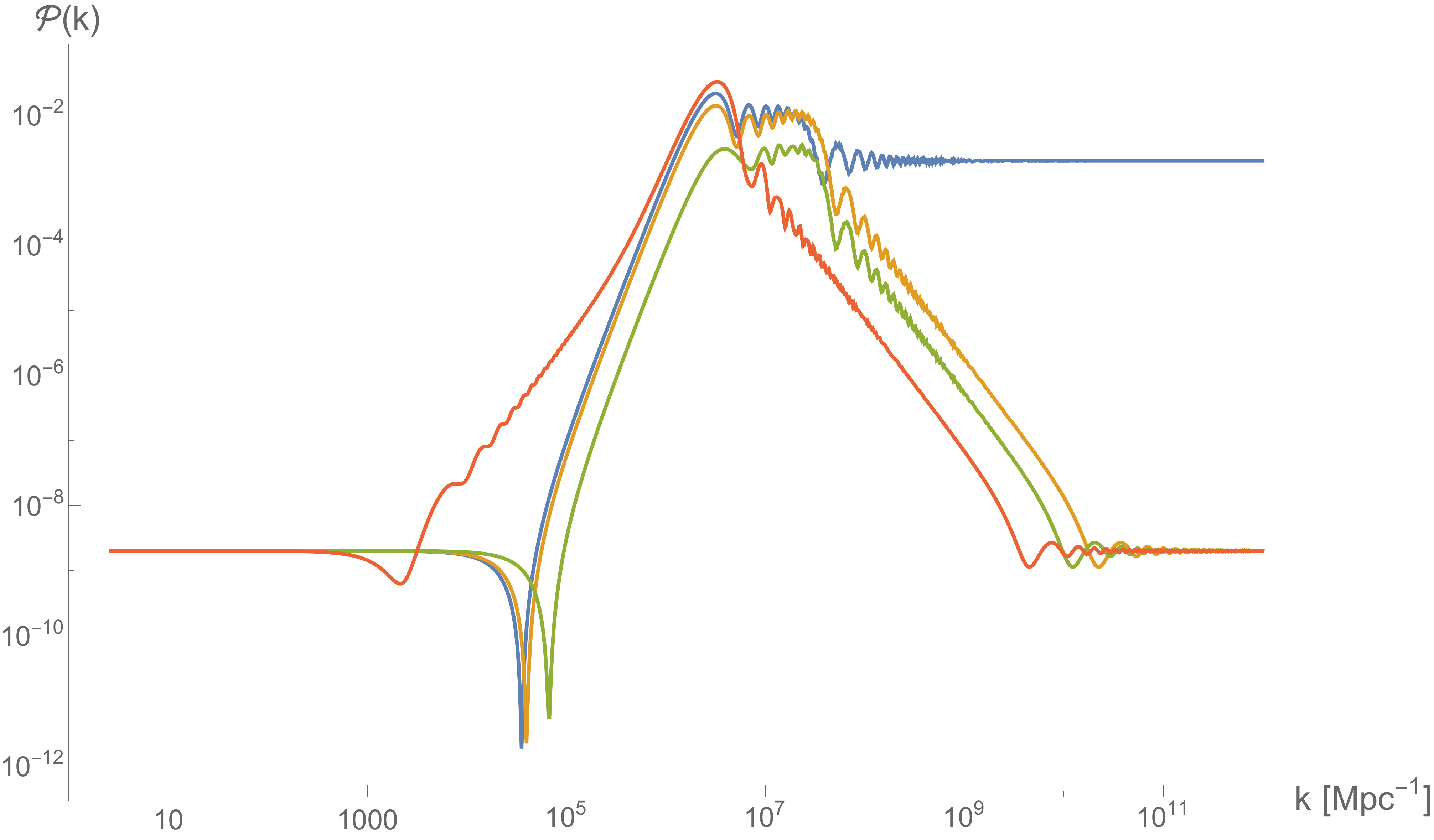}
\caption{Four power spectra involving different matchings between constant $\eta$ phases. The blue line is the same as the blue line plotted in Fig.~\ref{fig:different-etas}, matching from $\eta=0$ to $\eta=-6$ for 2.3 e-folds and back to $\eta=0$. The yellow line is a matching from $\eta=0$ to $\eta=-6$, then to $\eta=2$ and back to $\eta=0$. Notice that the peak amplitude decreases slightly when the positive $\eta$ phase is included - we comment on this further in App.~\ref{appendix:peakamp}. The green line is a 5-phase matching from $\eta=0$ to $\eta=-2$, then $\eta=-6$, then $\eta=2$ and back to $\eta=0$. The $\eta=-2$ phase does not decrease the slope of the power spectrum because the phase of $\eta=-6$ affects the scales that exit before the onset of the $\eta=-2$ phase, however it does cause the dip to occur at a larger value of $k$, and for the peak amplitude to be reduced. The red line is the same set-up as for the green line, but with a longer duration of $\eta=-2$ and shorter duration of $\eta=-6$ so that the $k^2$ growth is visible before the onset of the $k^4$ spectrum due to USR.}
\label{fig:4_matchings}
\end{figure}

We match the mode functions given by equation (\ref{msv}) for each of the five phases of constant $\eta$, at four successive transition times, and the final power spectrum is given by:

\eq{4match}{\calP_\calR = \lim_{\tau \to 0^-}\frac{k^3}{2\pi^2}|\calR_k^{(3)}|^2 =  \frac{H^2}{8\pi^2\mpl^2\epsilon_5}\left[c_4^*c_4 + s_4^*s_4- s_4^*c_4 - s_4c_4^*\right],}
with
\eq{eps5}{\epsilon_5= \epsilon_1[(a_3/a_2)^3(a_3/a_4)(a_2/a_1)]^{-2}=\epsilon_1[(\tau_2/\tau_3)^3(\tau_4/\tau_3)(\tau_1/\tau_2)]^{-2},}
and the coefficients $c_4$ and $s_4$ given by:

\tiny\eq{c4}{c_4=-\frac{i e^{i k \tau_4}}{16 k^9
   \tau_1 \tau_2^2 \tau_3^4
   \tau_4^2} \left(\begin{split} & (2 k
   \tau_1-i) (3+2 i k \tau_2)
   \left(2 k^4 \tau_3^4+4 k^2
   \tau_3^2+9\right) (k \tau_4+2 i)
   e^{-i k (2 \tau_2-\tau_4)}\\
   & +(2 k
   \tau_2 (2+i k \tau_2)-3 i)
   \left(2 k^4 \tau_3^4+4 k^2
   \tau_3^2+9\right) (k \tau_4+2 i)
   e^{-i k (2 \tau_1-\tau_4)}\\
   & +(2 k
   \tau_2+3 i) \left(2 k^4
   \tau_3^4+4 k^2 \tau_3^2+9\right)
   (k \tau_4 (3+2 i k \tau_4)-2 i)
   e^{-i k (2 \tau_1-2
   \tau_2+\tau_4)}\\
   & +(1+2 i k
   \tau_1) (-3+2 k \tau_2 (k
   \tau_2+2 i)) \left(2 k^4
   \tau_3^4+4 k^2 \tau_3^2+9\right)
   e^{-i k \tau_4} (-2+k \tau_4 (2 k
   \tau_4-3 i))\\
   & +(2 k \tau_2+3 i)
   (9+2 k \tau_3 (k \tau_3 (-7-2 i k
   \tau_3)+9 i)) (k \tau_4+2 i)
   e^{-i k (2
   (\tau_1-\tau_2+\tau_3)-\tau_4)}\\
   & +(2 k \tau_1-i) (-3+2 k
   \tau_2 (k \tau_2+2 i)) (9+2 k
   \tau_3 (k \tau_3 (-7-2 i k
   \tau_3)+9 i)) (k \tau_4+2 i)
   e^{-i k (2 \tau_3-\tau_4)}\\
   & +(2 k
   \tau_1-i) (2 k \tau_2-3 i) (3-2 i
   k \tau_3) (-3+2 k \tau_3 (k
   \tau_3+2 i)) (-2+k \tau_4 (2 k
   \tau_4-3 i)) e^{-i k (2 \tau_2-2
   \tau_3+\tau_4)}\\
   & +(-3+2 k \tau_2
   (k \tau_2-2 i)) (3-2 i k \tau_3)
   (-3+2 k \tau_3 (k \tau_3+2 i))
   (-2+k \tau_4 (2 k \tau_4-3 i))
   e^{-i k (2 \tau_1-2
   \tau_3+\tau_4)}\end{split}\right)}
   
\eq{s4}{s_4=\frac{e^{-2 i k (4
   \tau_1+\tau_2+\tau_3+\tau_4
   )}} {16 k^9 \tau_1 \tau_2^2
   \tau_3^4 \tau_4^2} \left(\begin{split}&(3-2 i k \tau_2) \left(2
   k^4 \tau_3^4+4 k^2
   \tau_3^2+9\right) (k \tau_4-2 i)
   e^{2 i k (3 \tau_1+2
   \tau_2+\tau_3)}\\
   & +(2 k \tau_1-i)
   (2 k \tau_2 (2-i k \tau_2)+3 i)
   \left(2 k^4 \tau_3^4+4 k^2
   \tau_3^2+9\right) (k \tau_4-2 i)
   e^{2 i k (4
   \tau_1+\tau_2+\tau_3)}\\
   & +(2 k
   \tau_1-i) (2 k \tau_2-3 i)
   \left(2 k^4 \tau_3^4+4 k^2
   \tau_3^2+9\right) (k \tau_4 (3-2
   i k \tau_4)+2 i) e^{2 i k (4
   \tau_1+\tau_3+\tau_4)}\\
   & +(-3+2 k
   \tau_2 (k \tau_2-2 i)) \left(2
   k^4 \tau_3^4+4 k^2
   \tau_3^2+9\right) (k \tau_4 (3-2
   i k \tau_4)+2 i) e^{2 i k (3
   \tau_1+\tau_2+\tau_3+\tau_4
   )}\\
   & +(-3+2 k \tau_2 (k \tau_2-2 i))
   (2 k \tau_3+3 i) (-3+2 k \tau_3
   (k \tau_3+2 i)) (2+i k \tau_4)
   e^{2 i k (3 \tau_1+\tau_2+2
   \tau_3)}\\
   & +(2 k \tau_1-i) (3+2 i k
   \tau_2) (2 k \tau_3+3 i) (-3+2 k
   \tau_3 (k \tau_3+2 i)) (k
   \tau_4-2 i) e^{4 i k (2
   \tau_1+\tau_3)}\\& +(2 k \tau_2+3
   i) (2 k \tau_3-3 i) (2 k \tau_3
   (2+i k \tau_3)-3 i) (-2+k \tau_4
   (2 k \tau_4+3 i)) e^{2 i k (3
   \tau_1+2 \tau_2+\tau_4)}\\
   & -(2 k
   \tau_1-i) (3-2 k \tau_2 (k
   \tau_2+2 i)) (3+2 i k \tau_3)
   (3-2 k \tau_3 (k \tau_3-2 i))
   (2-k \tau_4 (2 k \tau_4+3 i))
   e^{2 i k (4
   \tau_1+\tau_2+\tau_4)}\end{split}\right)}
\normalsize
The late-time power spectrum is shown in Fig.~\ref{fig:4_matchings}. The phase of $\eta=+2$ causes a decrease in power for large $k$, which we have chosen to return to the small-$k$ amplitude of $2\times10^{-9}$ for the red, yellow and green lines in Fig.~\ref{fig:4_matchings}, rather than the scale-invariant spectrum produced by matching straight back to $\eta=0$ as in previous sections, and shown by the blue line in Fig.~\ref{fig:4_matchings}. Also notice that the effect of the $\eta=-2$ phase is only visible if it lasts considerably longer than the phase of $\eta=-6$, otherwise the $k^4$ growth is dominant on the scales that the $\eta=-2$ phase affects.

\subsection{Peak amplitude sensitivity to late times}\label{appendix:peakamp}
\begin{figure}
\centering
\includegraphics[width=0.8\textwidth]{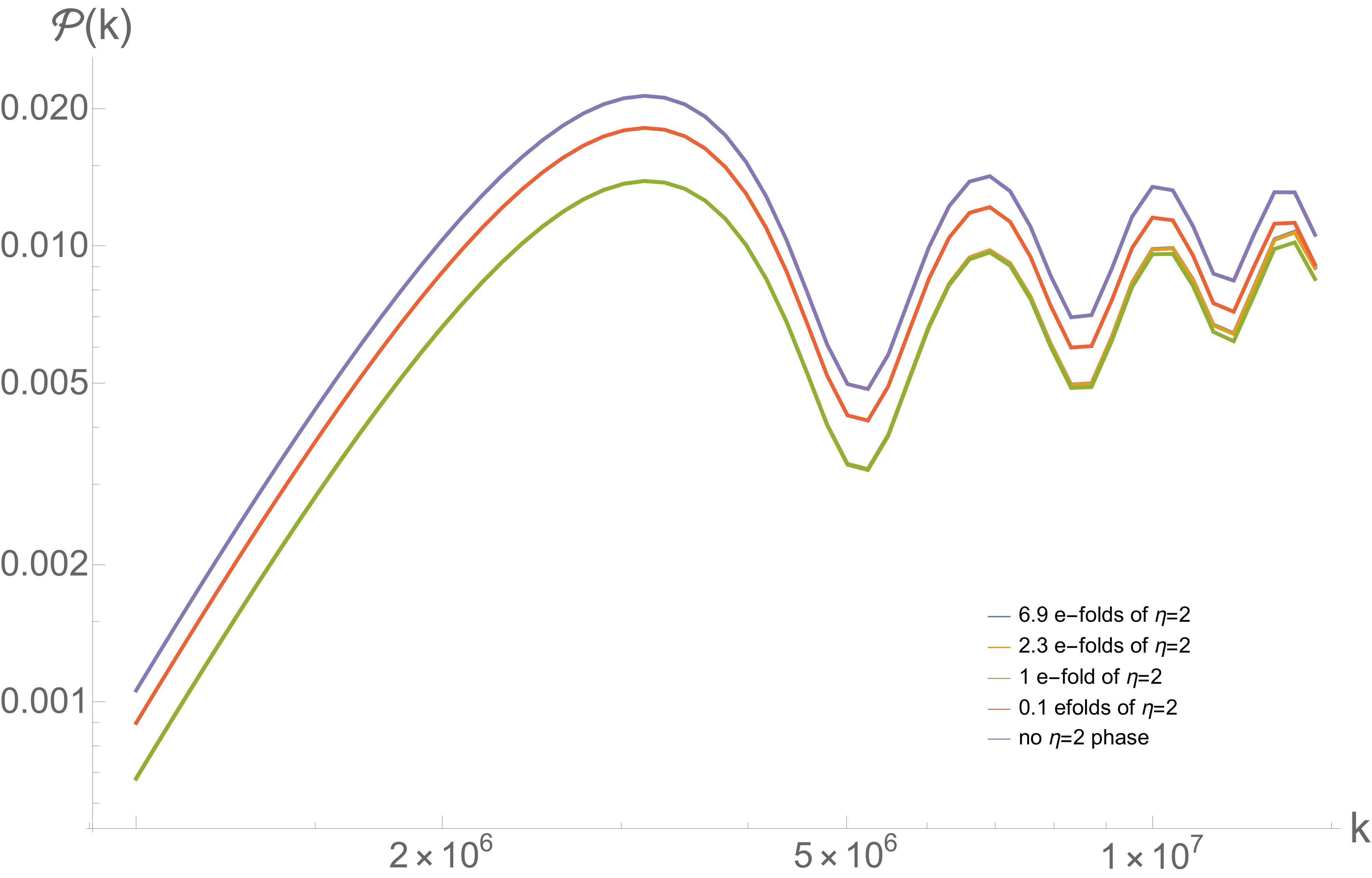}
\caption{Analytical power spectra with 4 phases of constant $\eta$: 0, -6, 2, 0. The only difference between the lines is the duration of the $\eta=2$ phase. The longer the phase of $\eta=2$, the less power at the very peak of the power spectrum, showing that how ultra slow roll ends has an effect on the amplitude of the peak. Notice that the spectra quickly converge to the amplitude for longer than 1 e-fold of $\eta=2$, the blue and yellow lines are hidden beneath the green line.}
\label{fig:peak_power}
\end{figure}

The amplitude of the peak of the power spectrum depends on how ultra slow roll finishes. How $\eta$ transitions back to 0 from a phase of $\eta=-6$ can shave off power from the peak. For example, if we set $\tau_1=\tau_2$ in the matching calculation from section \ref{sect:-2}, then we can plot the power spectrum for constant phases of $\eta$ from 0 $\rightarrow$ -6 $\rightarrow$ 2 $\rightarrow$ 0 so as to focus on the transition out of USR. In Fig.~\ref{fig:peak_power} the power spectrum is plotted for 6 different durations of $\eta=2$ -- all other parameters are kept the same -- with the different spectra being normalised such that the large scale amplitude is $2\times10^{-9}$. There is almost a factor of 2 difference in the peak amplitude between no $\eta=2$ phase and 1 e-folding of $\eta=2$ following ultra slow roll. However, the amplitude of the power spectrum is unaffected any further by increasing the duration of the $\eta=2$ phase beyond 1 e-folding.  While this is unlikely to have significance in terms of avoiding power spectrum constraints, it may have a large effect on the predicted number of PBHs produced, since the mass fraction is exponentially sensitive to the amplitude of the power spectrum. 
Note that this is for a sharp transition in $\eta$, and the effect may not be present for a smooth transition. This was investigated for the bispectrum in \cite{Cai:2017bxr}, where it was found that local non-Gaussianity is erased during a smooth exit from ultra slow roll, but that it can survive a sharp transition.

\section{The steepest constant $\eta$ spectrum}\label{section:k3}
If we consider the toy situation of an epoch of inflation defined by a constant, but non-zero $\eta < 0$ without matching to another epoch of inflation, we can arrive at simple bounds on how fast the power spectrum can grow given a constant $\nu$ and $\epsilon\ll1$. In this case, the general solution to the mode function corresponding to the Bunch-Davies vacuum is given by \eqref{msv}. If the late-time limit is taken directly without matching to any other phases, then
\be \P_\R\propto k^{3-2\nu},\ee
and the spectral index is given by
\be n_s-1=3-|3+\eta| \label{eq:ns-constant-eta},\ee
which gives a scale-invariant spectrum for $\eta=0,-6$ and the strongest possible positive scaling is $n_s-1=3$ for $\eta=-3$.

The steepest possible growth follows from setting $\eta=-3\Leftrightarrow\nu=0$ but because both modes are important in this case, the approximation \eqref{eq:ns-constant-eta} overestimates the actual slope, with the complete late time solution being 
\bea \P_\R&\propto& k^3\left(1+\frac{4}{\pi^2}\left(\gamma+\ln\left(\frac{k}{2 k_e}\right)\right)^2 \right), 
\label{eq:k3}\\ 
n_s-1&=& 3+\frac{8 \left(\log \left(\frac{k}{2 k_e}\right)+\gamma \right)}{\pi ^2 \left(\frac{4 \left(\log \left(\frac{k}{2 k_e}\right)+\gamma \right)^2}{\pi ^2}+1\right)}, \eea
where $\gamma=0.5772$ and $k_e=(a H)_e$ is the value of $k$ when this period of inflation at the boundary of USR ends (meaning that the decaying and growing modes are both important), assuming that the curvature perturbation freezes out afterwards. The correction to the $k^3$ scaling in \eqref{eq:k3} comes from the ``decaying'' mode which scales as
\be \R_{\rm decaying}\sim \int\frac{dN}{a^3 \epsilon} \propto \int dN\sim N =\log(k/k_e). \ee
The solution \eqref{eq:k3} agrees with \cite{Martin:2012pe} in the limit $\ln(k/k_e)\gg1$. The potential giving rise to this growth rate of the power spectrum in the limit $\epsilon\rightarrow0$, is
\be V=M^4 e^{\frac38\frac{\phi^2}{M_{Pl}^2}}. \ee
which can be calculated by using $\phi'\propto\sqrt{\epsilon}\propto e^{-3N}$ and the equation of motion \eqref{eq:KG}.

However, (\ref{eq:k3}) implies a weaker bound than the $k^4$ steepest growth index for single-field inflation shown via a more realistic matching calculation in App.~\ref{appendix:matching}. A complementary perspective is obtained by reconsidering the what a power spectrum with a constant growth index implies in position space. In order to do this, we consider the following form for the power spectrum
\eq{}{\calP_\calR \propto k^n e^{-\alpha k},}
which needs to be regulated for certain values of the index by a non-zero $\alpha$ which we take to zero at the end of the calculation. We recall that the position space two-point function of the curvature perturbation at late times is given by Fourier transform 
\eq{}{\int \frac{d^3k}{\left(2\pi\right)^3}e^{i\vec k\cdot\vec x }|\calR_k(0)|^2 = \langle\calR(\vec x,0)\calR(0,0)\rangle }
where 
\eq{}{\calP_\calR = \lim_{\tau \to 0^-}\frac{k^3}{2\pi^2}|\calR_k(\tau)|^2,}
and therefore
\eq{}{\langle\calR(\vec x,0)\calR(0,0)\rangle \propto \lim_{\alpha\to 0}\int \frac{d^3k}{4\pi}e^{i\vec k\cdot\vec x }\,k^{n-3}e^{-\alpha k}. }
For $n = 0$ we recover the usual logarithmically divergent position space correlation function (an artefact of us working in the strict dS limit). For $n>3$ we find
\eq{div}{\langle\calR(\vec x,0)\calR(0,0)\rangle \propto \lim_{\alpha\to 0}\int \frac{d^3k}{4\pi}e^{i\vec k\cdot\vec x }\,k^{n-3}e^{-\alpha k} \propto \frac{1}{|x|^n}. }
Therefore, asking why a power spectrum with a constant index $n$ can't have an index greater than $n = 4$ is the same as asking why the position space two-point function for the curvature perturbation can't diverge in the coincident limit faster than the fourth power of the distance between the two operators. The reason for this boils down to dimensional analysis. In a mass dependent regularisation scheme (i.e.~regulating divergences with a hard cut-off $\Lambda$), the more divergent a correlation function is in position space, the greater the power of divergence in momentum space. Two-point functions that diverge as the inverse square of the distance require counterterms proportional to $\Lambda^2$. Since our theory has no other UV mass scale, one cannot have a dependence on the r.h.s. of (\ref{div}) where $n$ is greater than 4, since this would require a counterterm that goes as $\Lambda^{n > 4}$, which is not possible in four dimensions. However, this does not completely account for the steepest growth shown in App.~\ref{appendix:matching}, since for any finite $\alpha$, the spectrum cuts off and the corresponding divergence is automatically regulated, invalidating the above argument. Although causality and analyticity arguments have been invoked in different contexts to argue for a particular bounds on the growth index of various cosmological perturbations\footnote{In the context of density perturbations produced from a causal collapse process in a non-inflationary context, Traschen and Abbott have derived a \textit{minimum} growth index of $k^4$ \cite{Abbott:1985di}. For primordial magnetic fields, Durrer and Caprini have shown that the two-point function must scale at least as $k^2$ at large scales \cite{Durrer:2003ja}.}, none of these appear to apply to our present context. The physical origin of the steepest growth index over a \textit{finite} range of modes that we've uncovered is still something we're investigating.

\section{On the background potential}\label{appendix:reconst}
In the first part of this appendix, we show how one can reconstruct a potential having specified an arbitrary time-dependence for $\epsilon$. Note that this is a much simpler problem than reconstructing the inflaton potential (more generally, action) from CMB data, a process that is necessarily hamstrung by a variety of degeneracies \cite{Kadota:2005hv,Cline:2006db,Bean:2008ga}. Our goal is simply to show that one can in principle design a potential assuming a minimally coupled scalar field with a canonical kinetic term to reproduce an arbitrary time-dependent profile for $\epsilon$. In the second part of this appendix, we show how one cannot engineer an arbitrarily abrupt end to inflation in terms of e-folds without introducing additional hierarchies that will be radiatively unstable. 

\subsection{Reconstructing $V$ from $\epsilon$}

We begin with the equation of motion for a minimally coupled scalar $\phi$, switching to  e-folding number $\calN$ as the time variable
\eq{neom}{H^2 \frac{d^2 \phi}{d\calN^2} + \left(3H^2 + H \frac{d H }{d\calN}\right)\frac{d \phi}{d\calN} + \frac{\partial V}{\partial \phi} = 0.}
Given that $H \frac{d H}{d\calN} = \dot H$, one can use the Friedmann equations $3H^2 = \rho$, $\dot H = -(\rho + p)/2\mpl^2$ to obtain
\eq{}{3H^2 + H \frac{d H}{d\calN} = 3H^2 + \dot H = \frac{\rho - p}{2\mpl^2} = \frac{V}{\mpl^2}.}
Furthermore the Einstein constraint equation becomes
\eq{hpp}{H^2\left(3 - \epsilon\right) = \frac{V}{\mpl^2}.}
Inserting these relations into (\ref{neom}) results in the equation of motion
\eq{eom}{\frac{d^2\phi }{d\calN^2} + \left[\frac{d \phi}{d\calN} + \frac{\mpl^2}{V}\frac{\partial V}{\partial\phi}\right]\left(3 - \epsilon\right) = 0  }
or
\eq{epseom}{\frac{d \epsilon}{d \calN} = -\left(3 - \epsilon\right)\left[2\epsilon +  \frac{d \phi}{d\calN}\frac{\partial_\phi V}{V}\right]}
where we have used $\epsilon = \frac{\left(d \phi/d\calN\right)^2}{2\mpl^2}$. So far, the above relations are exact. We now presume that $\epsilon \ll 3$ so that the above can be approximated as\footnote{Note that one can straightforwardly generalise the above derivation to the case of multi-field inflation, where the final equation (\ref{epseom1}) would also result. 
}
\eq{epseom1}{\frac{d \epsilon}{d \calN} = -6\epsilon + \frac{d \log V^{-3}}{d \calN}.}

Using the definition of $\epsilon$ and (\ref{epseom1}) we find
\eq{phieom}{\phi(\calN) = \phi_* \pm  \mpl\int^{\calN}_{\calN_*} d\calN' \sqrt{2\epsilon(\calN')},}
\eq{epseom2}{V(\calN) = V(\calN_*)\exp\left[ -\frac{1}{3}\int^{\calN}_{\calN_*} d\calN' \left(\frac{d \epsilon}{d \calN'} + 6\epsilon\right)     \right],}
giving us $\phi$ and $V$ as functions of $\calN$ determined entirely by the evolution of $\epsilon$ that we take as an input. It remains to figure out what $V$ is as a function of $\phi$. To do this, we observe that if
\eq{}{V(\calN) = \sum_{n=0} c_nf_n[\phi(\calN)]}
where the $f_n$ are some complete basis of functions\footnote{e.g. $f_n = \phi^n$ or $f_n = e^{n\lambda\phi}$ for some fixed $\lambda$ etc. In general, the convergence of the reconstructed potential to the true potential will depend greatly on choice of basis functions adopted, and the range in field space one wants the approximation to be valid.}, and if we know $V(\calN_i)$ and $\phi(\calN_i)$ for $0 \leq i \leq m$ discrete values, then if we demand than $V(\phi)$ truncate at some finite order $m$, we have a system of $m+1$ linear equations in $m+1$ unknowns which will be possible to invert given the presumption of monotonicity of $\phi$ and linear independence of the basis functions, allowing us to calculate the coefficients $c_i$ for $0 \leq i \leq m$, thus reconstructing an approximation to the potential to order $m$. For a limited enough field excursion it suffices to truncate to some small finite order e.g.~at $m = 6$ for a monomial basis: the typical order to which we need to know the potential in order to have a handle on the $\eta$ problem (see for instance, the treatment in \cite{Hertzberg:2017dkh}). 

However, for simple enough time dependence for $\epsilon$ one can explicitly perform a direct reconstruction. We first match a phase of constant $\epsilon$ slow roll to a phase of USR. First, note that using the definition $d\epsilon/d\calN = \epsilon\eta$, (\ref{epseom2}) can be recast as
\eq{epseom3}{V(\calN) = V_*\exp\left[ -\frac{1}{3}\int^{\calN}_{\calN_*} d\calN' \epsilon\left(\eta + 6\right)     \right].}
Therefore, it is clear that during USR, $V(\calN)$ remains constant as inflation progresses. Furthermore, during constant $\epsilon$ slow roll, $\epsilon(\calN) \equiv \epsilon_0$ and $\eta = 0$, so that during this phase
\eq{}{V(\calN) = V_* e^{-2\epsilon_0(\calN - \calN_*)}.}
Picking the $+$ branch of the solution (\ref{phieom})
\eq{phi0}{\phi(\calN) - \phi_* = \mpl\sqrt{2\epsilon_0}(\calN - \calN_*)}
we find that we can straightforwardly invert $\phi$ for $\calN$, resulting in the potential
\eq{}{V(\phi) = V_*e^{-\frac{\sqrt{2\epsilon_0}}{\mpl}(\phi - \phi_*)},}
which is consistent with the fact that the only constant $\epsilon$ attractors are given by exponential potentials. Next, we note that during USR, the argument of (\ref{epseom3}) vanishes identically, so that the potential during this phase has a constant value set by the value at the end of the constant SR epoch --
\eq{}{V(\calN) = V_*e^{-2\epsilon_0(\calN_1 - \calN_*)} = {\rm const.}~~~ \calN> \calN_1.}
Similarly, given that $\epsilon(\calN) = \epsilon_0e^{-6(\calN - \calN_1)}$ during USR, we find from (\ref{phieom}) that
\eq{}{\frac{3(\phi - \phi_1)}{\sqrt{2\epsilon_0}\mpl} = 1 - e^{-3(\calN - \calN_1)}}
with $\phi_1$ given by (\ref{phi0}) evaluated at $\calN_1$. The only way some polynomial function of the above can result in a constant is if it were itself a constant. Hence the reconstructed potential that transitions from slow roll to USR is a piecewise potential that glues an exponential potential to a constant. This is not particularly physical, so we can try to suitably smooth the transition from SR to USR. 

We now reproduce a potential that can mimic the matching calculation done in the previous appendix. Namely, from $\eta = 0$ slow roll to $\eta = -2 \to \eta = -6 \to \eta = +2$ back to $\eta = 0$ slow roll. When $\eta = 0 \to \eta = -2$ at $\calN = \calN_1$, we can repeat the steps above for $\calN > \calN_1$ to find
\eq{v2}{V(\calN) = V(\calN_1) e^{\frac{2\epsilon_0}{3}\left[e^{-2(\calN-\calN_1)} - 1 \right] } }
and similarly for the field profile,
\eq{p2}{\frac{\phi - \phi_1}{\mpl \sqrt{2\epsilon_0}} = 1 - e^{-(\calN - \calN_1)}.}
Substituting the above into the exponent of (\ref{v2}) results in
\eq{}{V(\phi) = V(\phi_1)e^{\frac{(\phi - \phi_1)^2}{3\mpl^2} - \sqrt{2\epsilon_0}\frac{2(\phi - \phi_1)}{3\mpl} }.}
Note that from (\ref{p2}) the field can only asymptote to $\phi - \phi_1 = \sqrt{2\epsilon_0}\mpl$, in which case the potential goes to zero smoothly. From the previous discussion, we see that to match to $\eta = -6$ is to splice this potential to a constant piece at $\calN = \calN_2$. To subsequently match from this phase to $\eta = +2$ at $\calN_3$ results in (for $\calN > \calN_3$)
\eq{v3}{V(\calN) = V(\calN_3) e^{-\frac{4\epsilon_2}{3}\left[e^{2(\calN-\calN_3)} - 1 \right] } }
during which time the field evolves as
\eq{}{\phi - \phi_3 = \mpl\sqrt{2\epsilon_2}\left[e^{\calN-\calN_3} - 1\right] }
so that the potential this corresponds to is given by
\eq{}{V(\phi) = V(\phi_3)e^{\frac{-2(\phi - \phi_3)^2}{3\mpl^2} - \sqrt{2\epsilon_2}\frac{4(\phi - \phi_3)}{3\mpl} } }
with 
\eq{e2}{\epsilon_2 = \epsilon_0 e^{-2(\calN_2 - \calN_1)}e^{-6(\calN_3 - \calN_2)}.}
Finally, one would like to match to a slow roll phase again, where
\eq{}{V(\phi) = V(\phi_4)e^{-\frac{\sqrt{2\epsilon_3}}{\mpl}(\phi - \phi_4)}}
with 
\eq{e3}{\epsilon_3 = \epsilon_0 e^{-2(\calN_2 - \calN_1)}e^{-6(\calN_3 - \calN_2)}e^{2(\calN_4 - \calN_3)}}
Therefore we summarize that the piecewise continuous potential that reproduces the matching $\eta = 0 \to \eta = -2 \to \eta = -6 \to \eta = +2 \to \eta = 0$ is given by
\begin{eqnarray}
\label{recon}
V_1(\phi) &=& V_*e^{-\frac{\sqrt{2\epsilon_0}}{\mpl}(\phi - \phi_*)}~~~~~~~~~~~~~~~~~~~~~~ \phi < \phi_1, ~\eta = 0\\
\nn V_2(\phi) &=& V_1(\phi_1)e^{\frac{(\phi - \phi_1)^2}{3\mpl^2} - \sqrt{2\epsilon_0}\frac{2(\phi - \phi_1)}{3\mpl} }~~~~~~~ \phi_1 < \phi < \phi_2,~ \eta = -2\\
\nn V_3(\phi) &=& V_2(\phi_2) = {\rm constant}~~~~~~~~~~~~~~~~~ \phi_2 < \phi < \phi_3,~ \eta = -6\\
\nn V_4(\phi) &=& V_3e^{\frac{-2(\phi - \phi_3)^2}{3\mpl^2} - \sqrt{2\epsilon_2}\frac{4(\phi - \phi_3)}{3\mpl} }~~~~~~~~~~ \phi_3 < \phi < \phi_4,~ \eta = +2 \\
\nn V_5(\phi) &=& V_4(\phi_4)e^{-\frac{\sqrt{2\epsilon_3}}{\mpl}(\phi - \phi_4)}~~~~~~~~~~~~~~~~ \phi_4 < \phi, ~\eta = 0
\end{eqnarray}
with $\epsilon_{2}$ and $\epsilon_{3}$ given by (\ref{e2}) and (\ref{e3}), and where the fixed field intervals  in terms of the number of e-folds of the different phases as
\begin{eqnarray}
\phi_2 - \phi_1 &=& \mpl \sqrt{2\epsilon_0}\left[1 - e^{-(\calN_2 - \calN_1)}\right]\\ \nn
\phi_3 - \phi_2 &=& \frac{\mpl}{3} \sqrt{2\epsilon_0}e^{-(\calN_2- \calN_1)}\left[1 - e^{-3(\calN_3 - \calN_2)}\right]\\ \nn
\phi_4 - \phi_3 &=& \mpl \sqrt{2\epsilon_0}e^{-(\calN_2- \calN_1)}e^{-3(\calN_3- \calN_2)}\left[e^{(\calN_4 - \calN_3)} - 1\right].
\end{eqnarray}
We plot the reconstructed potential below for specific values of the $\calN_i$:
\begin{figure}[H]
\centering
\includegraphics[width=10cm]{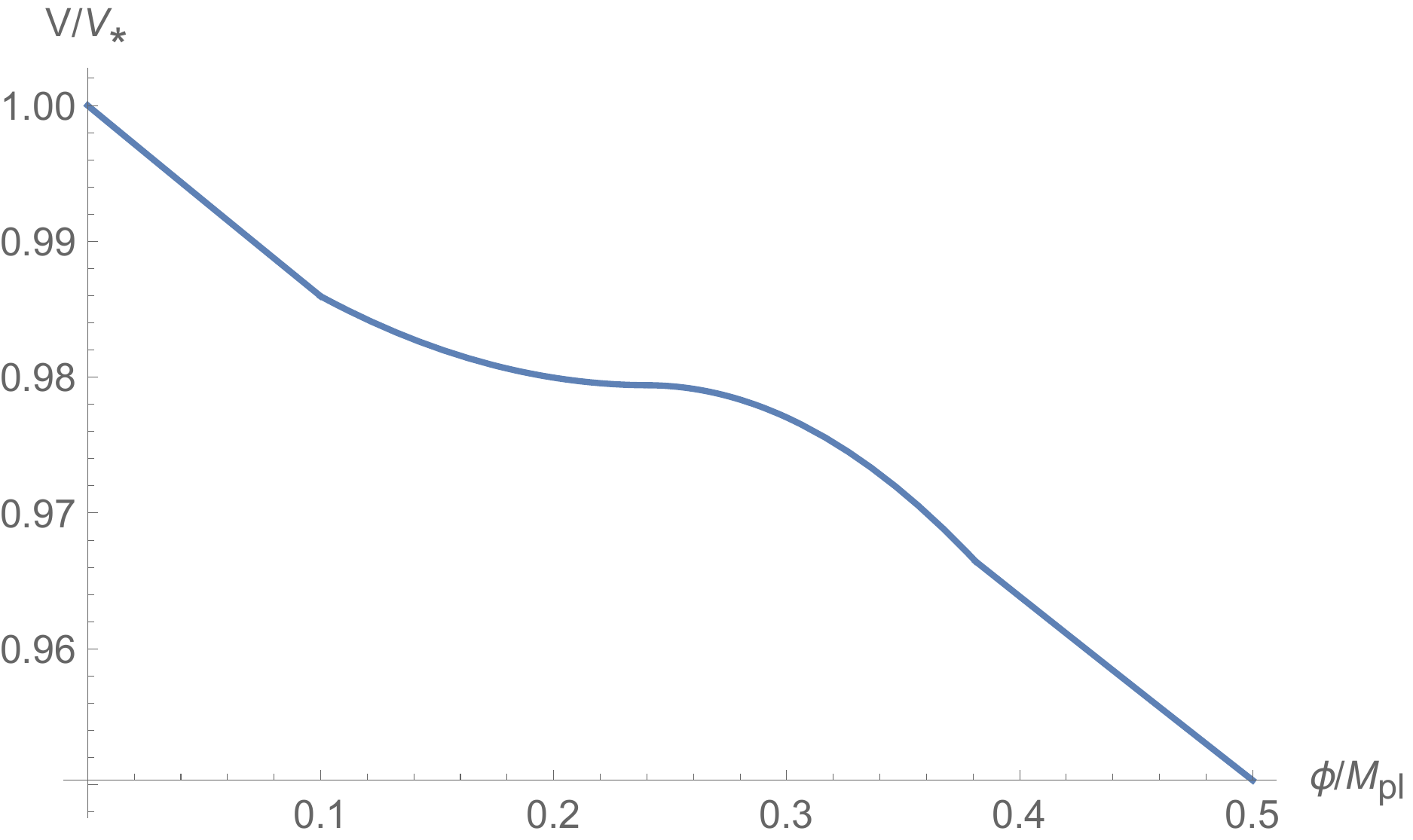}
\caption{The reconstructed potential (\ref{recon}) for $\calN_1, 
\calN_2, \calN_3, \calN_4 = 10, 14, 18, 34$ respectively, with $\phi_* = 0$ and $\epsilon_0=0.01$. Note that the field range over which USR occurs ($\phi_3 - \phi_2\simeq0.0009\mpl$) is parametrically much smaller than the phases where $\eta = \pm 2$, so as to effectively appear as an inflection point in the above plot.}
\label{reconstructed_V}
\end{figure}

\subsection{The quickest possible end to inflation}\label{sec:end}
Having demonstrated a steepest possible growth for the primordial power spectrum, one might wonder about the complementary question -- how quickly can it fall off? An accurate estimate for this can be inferred from rephrasing the question as how quickly inflation can end, or transition to another phase of inflation. To understand this, we first recall (\ref{eom}), but now generalise to multi-field inflation

\eq{eommf}{ \frac{d^2\phi^a}{d\calN^2} + \left[\frac{d\phi^a}{d\calN} + \frac{\mpl^2}{V}\frac{\partial V}{\partial\phi_a}\right]\left(3 - \epsilon\right) = 0  }
where $\phi^a$ denotes coordinates in some general field space. For simplicity, we assume a flat field space metric (and so accord no significance to raised or lowered indices) although this can be straightforwardly generalised. The slow-roll parameter $\epsilon$ is now defined as
\eq{}{\epsilon =  \frac{1}{2\mpl^2}\frac{d\phi^a}{d\calN}\frac{d\phi_a}{d\calN}.}
Multiplying (\ref{eommf}) by $d\phi_a/d\calN$ results in the analog of (\ref{epseom}) 
\eq{epseomf}{\frac{d \epsilon}{d \calN} = -\left(3 - \epsilon\right)\left[2\epsilon +  \frac{d \phi^a}{d\calN}\frac{\partial_aV}{V}\right].}
We now consider the situation where over some interval, $\epsilon$ increases monotonically from some initial $\epsilon_0 \ll 1$ to $\epsilon_f = 1$ over an interval $\Delta\calN_{\rm end}$. Application of the mean value theorem of calculus\footnote{Recalling that if $f$ and $f'$ are continuous functions on the interval $[a,b]$, then there exists some point $c \in [a,b]$ such that $f'(c) = \frac{f(b) - f(a)}{b-a}$. Since $f'$ is also continuous, $f'(c)$ must lie between the minimum and maximum of $f'$ in the interval $[a,b]$. That is \eq{}{\nn \min_{a \leq x \leq b} f'(x) \leq  \frac{f(b) - f(a)}{b-a} = f'(c) \leq  \max_{a \leq x \leq b} f'(x).}} then implies that
\eq{}{\bigg|\frac{d\epsilon}{d\calN} \bigg|_{\rm int} \gtrsim \frac{1 - \epsilon_0}{\Delta\calN_{\rm end}}  \sim \frac{1}{\Delta\calN_{\rm end}} }
at some intermediate $\calN_{\rm int}$. Inserting the rhs of (\ref{epseomf}) into the above, assuming $\Delta \calN_{\rm end} \ll 1$ and applying the triangle inequality results, after some manipulation, in the lower bound 
\eq{mine}{\mpl\bigg| \frac{\nabla_TV}{V} \bigg|_{\rm max} \gtrsim \frac{1}{3\sqrt 2\Delta\calN_{\rm end}} }
where $\nabla_TV$ is the tangential derivative of the potential with respect to the trajectory, defined as $\nabla_T V :=   T^a\partial_a V$ and $T^a := \frac{d\phi^a}{d\calN}\Big/\left(\frac{d\phi^b}{d\calN}\frac{d\phi_b}{d\calN}  \right)^{1/2}$. In the single-field case, it reduces to the more familiar expression
\eq{}{\mpl\bigg| \frac{V_{,\phi}}{V} \bigg|_{\rm max} \gtrsim \frac{1}{3\sqrt 2\Delta\calN_{\rm end}}.}
Therefore, if we would like inflation to end in $\Delta\calN_{\rm end} \ll 1$ e-folds or less, we necessarily require the gradient of the potential along the trajectory as inflation ends to be bounded from below according to (\ref{mine}). Although classically we are entitled to make the transition out of inflation as sharp as we desire, one cannot make it arbitrarily sharp without introducing additional hierarchies that will be unstable under quantum corrections, since these corrections spoil the flatness of the potential away from the transition, in effect ending inflation earlier and restoring the smoothness of the transition. Nevertheless, from (\ref{mine}) we see that a transition that lasts an order unity fraction of an e-fold can easily be accommodated without introducing additional hierarchies, and for the purposes of our discussion, justifies any approximation that cuts off the primordial power spectrum at some fixed comoving scale. 

For completeness, we illustrate the considerations above with a concrete example. Consider the following prototype potential for a rapid exit from inflation
\eq{pot}{V(\phi) = \frac{V_*}{2}e^{-\gamma\phi/\mpl}\left(1 - {\rm Tanh}\left[\mu(\phi-\phi_*)/\mpl \right]\right).}
When $\gamma \ll 1$, one has power law inflation in the region $\phi \ll \phi_*$. At $\phi = \phi_*$, there is a transition (that can be made arbitrarily abrupt as the dimensionless parameter $\mu \to \infty$). Requiring the transition to last less than 1/100th of an e-fold requires for example $\mu$ to be at least of order $10^2$ through (\ref{mine}), which would imply that the hyperbolic tangent is an operator expansion in odd powers of effective operators with very large Wilson coefficients:
\eq{}{\calL \supset \frac{\mu^{n}\phi^{n}}{\mpl^{n}};~~~\mu \sim 10^2, ~\Delta\calN_{\rm end} \sim 10^{-2}.}
Calculating loop corrections to the potential (\ref{pot}) expanded around $\phi_*$ for $\mu \sim 10^2$ would result in a deformation of the inflationary part of the potential. If one were to try and approximate it close enough to $\phi_*$ as an exponential, one would find an effectively renormalised $\gamma$ that is no longer $\ll 1$. On the other hand, requiring $\Delta\calN_{\rm int} \sim 10^{-1}$ is possible for values of $\mu \sim 1$, resulting in a renormalisation group improved potential where the hierarchy $\gamma \ll 1$ is preserved. We stress however that the bound (\ref{mine}) is completely general and can be applied to multi-field inflation as well.

\bibliography{USR_bib}

\end{document}